\documentclass[10pt]{article}
\usepackage{amsmath}
\usepackage{amssymb}
\numberwithin{equation}{section}
\usepackage{amsthm}
\pdfoutput=1

\usepackage[utf8]{inputenc}
\usepackage[T1]{fontenc}
\usepackage{microtype}

\usepackage{fourier}
\usepackage{caption}
\usepackage{mathrsfs}
\usepackage[colorlinks=true, pdfstartview=FitV, linkcolor=blue, citecolor=blue, urlcolor=blue]{hyperref}
\theoremstyle{plain}
\newtheorem{proposition}{Proposition}[section]
\newtheorem{corollary}[proposition]{Corollary}
\newtheorem{lemma}[proposition]{Lemma}
\newtheorem{theorem}[proposition]{Theorem}
\theoremstyle{definition}
\newtheorem{definition}[proposition]{Definition}
\newtheorem{example}[proposition]{Example}
\newtheorem{remark}[proposition]{Remark}

\usepackage[usenames,dvipsnames]{xcolor}
\usepackage{url}
\usepackage{tikz}
\usepackage{pgfplots}
\pgfplotsset{compat=newest,ticks=none}
\usetikzlibrary{arrows,calc}
\usepackage{verbatim}
\usetikzlibrary{%
    decorations.pathreplacing,%
    decorations.pathmorphing%
}

\DeclareMathOperator{\sgn}{sgn}

\DeclareMathOperator{\sech}{sech}

\newcommand{\R}{\mathbf{R}}
\newcommand{\C}{\mathbf{C}}
\newcommand{\Oh}{\mathcal{O}}
\newcommand{\bb}{\omega}

\newcommand{\abs}[1]{\left\lvert #1 \right\rvert}

\newcommand{\In}{\mathcal{L}_{\rho}}
\newcommand{\Inm}{\mathcal{L}_{m}}
\usepackage{mathscinet}
\usepackage{cite}

\usepackage{microtype}
\begin{document}

\title{On isospectral deformations of an inhomogeneous string}

\author{
Kale Colville \thanks{
Department of Physics, McGill University, 3600 rue University, Montr\'{e}al, QC H3A 2T8, Canada; colvilk@physics.mcgill.ca}
\and 
  Daniel Gomez\thanks{Department of Mathematics, 
the University of British Columbia, 
1984 Mathematics Road, 
Vancouver,V6T 1Z2, Canada; dagubc@math.ubc.ca}
  \and
  Jacek Szmigielski\thanks{Department of Mathematics and Statistics, University of Saskatchewan, 106 Wiggins Road, Saskatoon, Saskatchewan, S7N 5E6, Canada; szmigiel@math.usask.ca}
}

\date{\today}

\maketitle

\begin{abstract}
  In this paper we consider a class of isospectral deformations of the inhomogeneous 
  string ~ boundary value problem.  The deformations considered 
  are generalizations of the isospectral deformation that has arisen in connection 
  with the Camassa-Holm equation for the shallow water waves.  It is proved that 
  these new isospectral deformations result in evolution equations on the mass density 
  whose form depends on how the string is tied at the endpoints.  Moreover, it is shown 
  that the evolution equations in this class linearize on the spectral side and hence 
  can be solved by the inverse spectral method.  In particular, the problem involving a mass density given by a discrete finite measure and arbitrary boundary conditions is shown to be solvable by Stieltjes' continued fractions.  
  
\end{abstract}

\tableofcontents

\section{Introduction}
The wave equation $\frac{1}{c^2} f_{\tau\tau} -\nabla ^2 f=0$ is a prototype of hyperbolic 
problems well studied since the time of Euler.  One of its many applications, 
well known to any science undergraduate student, is the one dimensional 
version of this equation $\frac{1}{c^2}f_{\tau\tau} -f_{xx}=0, \, 0<x<L$, describing 
the motion of a vibrating string of length $L$ with the amplitude $f(x, \tau)$ at 
the point $x$ and time $\tau$.  Usually, the equation is also equipped 
with boundary conditions expressing the way the endpoints of the string are behaving; 
the most common types are, of course, the case of the string with both ends fixed (i.e. $f(0,\tau)=f(L,\tau)=0$ ) or the case of the string with sliding endpoints (i.e. $f_x(0,\tau)=f_x(L,\tau)=0$).  The two cases are sometimes referred to as the Dirichlet and the Neumann strings respectively.  The coefficient $c^2$ has the physical dimension of the velocity squared but 
more appropriately it depends on the mass density of the medium.  In general, in the 
case of an inhomogeneous medium, this coefficient is position dependent.  
The string counterpart of this is an {\sl inhomogeneous string} equation: 
$\frac1{c^2(x)} f_{\tau \tau }-f_{xx}=0, \, 0<x<L $ with the position dependent speed $c^2(x)=\frac{T}{\rho(x)}$, where $T$ is the tension and $\rho$ 
	the lineal mass density of the string, 
	which leads, after separation of variables $f(x,\tau)=\cos(\sqrt{z}\tau)v(x)$, to 
	\begin{equation}\label{eq:istring}
		-v_{xx} = \frac zT\rho(x) v.  
	\end{equation} 
We will choose the tension $T=1$ and for simplicity of presentation  we will restrict our attention to the string problem 
with $L=1$.  The general string boundary value problem 
then reads: 
\begin{equation}\label{eq:stringBVP}
-v_{xx} =  z\rho(x) v, \qquad 0<x<1, 
\qquad v_x(0)-hv(0)=0, \quad v_x(1)+Hv(1)=0, 
\end{equation}
where, traditionally, $h=0$ corresponds to the Neumann boundary condition on the left end 
of the string, while $h=\infty$ corresponds to the Dirichlet condition on the left end, 
and the same convention applies to $H$ and the right end of the string. 
We will subsequently refer to equation \eqref{eq:stringBVP} as the string equation, 
although the name of the Helmholtz equation is also associated with this very equation.   

The mathematical pedigree of equation \eqref{eq:stringBVP} parallels its physical 
significance: \eqref{eq:stringBVP} impacted the development of 
the Fourier methods, continued fractions, and subsequently led to 
the fundamental progress associated with distributional (weak) solutions to partial differential equations.  
For this paper, however, 
the most relevant is M.G. Krein's pioneering work on the inverse problem 
for the inhomogeneous string\cite{Kreinstring, KK, Dym-McKean-Gaussian}.  
The inverse problem for the inhomogeneous string, in its simplest formulation, 
amounts to the reconstruction of the mass density $\rho$ from the spectrum of 
the boundary value problem.  It has been known since the time of Borg \cite{Borg} 
that the mass density of the string is not uniquely determined by the spectrum alone.  
It was M.G. Krein who shifted the emphasis to the reconstruction of the string density 
from the spectral function, sometimes also called the Weyl function.  In the simplest 
case of the Dirichlet spectrum the corresponding Weyl function is $W(z)=\frac{v_x(1;z)}{v(1;z)}$, where $v(x;z)$ satisfies the initial value problem: $-v_{xx} =  z\rho(x) v, \, 0<x<1, v(0;z)=0, v_x(0;z)=1$.  The zeros of the denominator of the Weyl function correspond 
to the spectrum of the Dirichlet boundary value problem, while the zeros of the numerator 
correspond to the spectrum of the boundary value problem with the Dirichlet condition on the left and 
the Neumann condition on the right.  
The inverse problem for the string appears in many different areas of science and engineering such as the geophysical inverse problems ~\cite{Barcilon-LNM} or magnetotelluric inversion ~\cite {Weidelt, HSS}, to mention just a few.  The reader might want to consult a very accessible account 
of inverse problems relevant to engineering in the book by G.M.L. Gladwell \cite{Gladwell}.

Since in general the knowledge of a single spectrum does not 
determine the string density, a natural question is to determine {\sl isospectral} deformations of the
string, i.e. changes in the density that leave invariant the spectrum of a given boundary value problem. 

Some results 
dealing with isospectral strings are presented in \cite{Gladwell} and also in the works 
of H.P.W Gottlieb \cite{Gottlieb} and in the references therein.  Our interest lies 
in   characterizing isospectral strings in terms of {\sl nonlinear evolution equations} 
on the mass density $\rho$.  The first contribution to this line of research, as far as we know, goes back to the 
series of interesting papers by P.C. Sabatier \cite{Sabatier-Evolution, Sabatier-Constants, Sabatier-NEEDS} (see Example \ref{ex:HD}).  Our approach, however, is more 
influenced by the later work on the Camassa-Holm equation (CH)  \cite{CH}: 
\begin{equation*}
m_t=m_x u +2u_x m,  \quad    m=u-u_{xx}.  
\end{equation*}
It was discovered in the late nineties in \cite{BSS-acoustic} that the CH equation 
can be viewed as an isospectral deformation of the inhomogeneous  string with Dirichlet boundary conditions.  
This led, among other things, to the construction of a certain class of explicit solutions (peakons) in \cite{BSS-Stieltjes} by 
adapting an elegant approach of Stieltjes \cite{Stieltjes}.  Some of these developments 
are reviewed in \cite{BSS-stringCH}.  

The present paper addresses the following questions: 
\begin{enumerate} 
\item consider a general class of deformations of the boundary value problem \eqref{eq:stringBVP} postulated to be described by $v_t=a v+bv_x$, 
where all functions depend on $x$, the deformation parameter $t$, and the spectral parameter $z$ appearing in equation \eqref{eq:stringBVP}.  Can
$a$ and $b$ be suitably chosen for the boundary value problem \eqref{eq:stringBVP} to remain isospectral?  
\item which of the isospectral nonlinear evolution equations on the mass density give rise to a linear evolution of the 
appropriate Weyl function?  
\end{enumerate}

In the body of the paper we give a definite, affirmative, answer to the first question 
under the assumption that $b$ is a rational function of $z$, regular 
at $\infty$ and possessing only simple poles which, for technical reasons, we 
assume to be located on the positive part of the real axis.

Even though the second question is less natural from the point of view of real life 
applications it  is  nevertheless of considerable mathematical interest because the linear evolution of the Weyl function leads, 
via the spectral/inverse scattering approach, to exact formulas which can be of great benefit, 
for example, in testing numerical methods. Furthermore, by constructing 
explicit solutions to nonlinear partial differential equations, one frequently 
gets valuable insight into the expected behaviour of a larger class of solutions 
after applying, for example, density arguments if such are available.  The research presented in this paper provides a partial answer to the question of linearization for the the following reason: we  rely on a particular scheme of linearization originally known from the inverse scattering approach to the KdV equation and subsequently applied to a multitude of integrable partial differentiable equations.  This method hinges
in an essential way on the linearization of dynamics in the asymptotic region in space, eventually leading to a linear evolution of the scattering data.  In summary, we answer the question of linearization only within the paradigm of the inverse scattering approach.

The paper is organized as follows: in section \ref{sec:deformations} we analyze what 
we call the Zakharov-Shabat deformations (ZS) modeled after those occurring in the 
theory of integrable systems (see, for example, \cite{FT}); section \ref{sec:BCs} contains a preliminary analysis 
of these flows, we derive a criterion for the preservation of boundary conditions under the deformations and provide evidence that the ZS isospectral deformations of the string arise from  
the class of flows analytic around 
$z=\infty$;  section \ref{sec:lin} contains, in Proposition \ref{prop:cond-isospectrality}, the description of these isospectral deformations in terms of  
evolution equations on the spectral data; in section \ref{sec:bplus} 
we prove that the deformation, rational in the spectral variable, regular 
at infinity, with simple poles at positive $\epsilon_1, \epsilon_2, \dots, \epsilon_M$, can be chosen to render equation 
\eqref{eq:stringBVP} not only isospectral but also such that the 
evolution of the spectral data is linear;  in  section \ref{sec:negativeflows} we 
show explicitly how the deformations work for discrete mass densities; 
finally, in Appendix \ref{app:weak formulation} we provide details behind 
the modifications of the  formalism when the mass densities are  not smooth but 
rather finite discrete measures, while in Appendix \ref{ap:mapping to R} we explain the main steps illustrating the relation of the string deformation equations with the CH equation.

\section{Inhomogeneous string and its deformations } \label{sec:deformations}

We will make the general assumption that $\rho$ is a positive measure and $h,H\geq 0$ to ensure that the spectrum of the boundary value problem \eqref{eq:stringBVP} is positive. However, to simplify the technical aspects of the paper, we will focus on two extreme cases: 
\begin{enumerate} 
\item $\rho$ is in $C^1([0,1])$, 
\item $\rho$ is a finite, discrete measure, i.e. $\rho=\sum_{j=1}^N m_j \delta_{x_j}$ where $\delta_{x_j}$ is the Dirac measure centered at 
$0<x_j<1$,  
\end{enumerate} leaving other scenarios of the smoothness of the mass 
density for future work.  
	
We begin by writing the string equation as 
 a first order system:
\begin{equation}\label{eq:xLax}
V_x=\begin{bmatrix} 0&1\\-z\rho &0 \end{bmatrix}V , \quad \text {where } V=\begin{bmatrix}v\\v_x\end{bmatrix},  
\end{equation}
and define deformations of the string equation as variations of 
the mass density $\rho$ expressed as a function of an external parameter $t$
which, we emphasize, is not a physical time, even though we will occasionally refer to it as time.  Our goal is to describe a class 
of deformations which leave invariant the spectrum of the string boundary value problem 
\eqref{eq:stringBVP}; we will refer to these deformations as \emph{isospectral deformations}  and will search for them among  deformations of the Zakharov-Shabat (ZS) type.  
\begin{definition}[ZS deformations]
\mbox{}\\*
$V$ in equation \eqref{eq:xLax}  is postulated to depend on the deformation parameter $t$ subject to 
\begin{equation}\label{eq:tLax}
V_t=\begin{bmatrix} a&b\\c&d \end{bmatrix}V
\end{equation}
where $a, b,c,d$ are functions of $x,t,z$, chosen so as to ensure   
the compatibility of equations \eqref{eq:xLax} and \eqref{eq:tLax}.

\end{definition}
 When $\rho, a, b,c,d$  are sufficiently smooth in $x$ and $t$ (for example all in $C^1$) the equations \eqref{eq:xLax} and \eqref{eq:tLax} are 
 compatible provided  the \emph{Zero Curvature condition} : 
\begin{equation}\label{eq:ZCC} 
\begin{bmatrix} 0&1\\-z\rho &0 \end{bmatrix}_t-\begin{bmatrix} a&b\\c&d \end{bmatrix}_x+
\big[\begin{bmatrix} 0&1\\-z\rho &0 \end{bmatrix},\begin{bmatrix} a&b\\c&d \end{bmatrix}\big]
=0, 
\end{equation}
holds, which, upon closer inspection, implies that 
 one can parametrize $a,c,d$ by the function $b$.  The result of a simple computation is
\begin{equation}\label{eq:abc}
        a = -\frac{1}{2}b_{x} + \beta, \qquad 
	c = -\frac{1}{2}b_{xx}-z\rho b, \qquad d = \frac{1}{2}b_{x} + \beta, 
\end{equation}
where $\beta$ is a constant (in $x$), whose choice, to be discussed later, is dictated by the boundary conditions.  
Then we substitute these into the equation containing $\rho_{t}$ to get
the following \emph{evolution deformation equation} 
\begin{equation}\label{eq:deformationeq} 
	z\rho_t =\frac{1}{2}b_{xxx}+z\rho_{x}b+2z\rho b_{x}, 
\end{equation}
which we will write as 
\begin{center} 
\boxed {z\rho_t =\frac{1}{2}b_{xxx}+z\mathcal{L}_{\rho}b}
\end{center}
with  
\begin{equation}\label{eq:Lrho}
  \In=\rho \frac{\partial}{\partial x} + \frac{\partial}{\partial x} \rho
\end{equation}
denoting the second and third term in \eqref{eq:deformationeq}.  
In general, one cannot say much more about equation ~\eqref{eq:deformationeq} unless some additional structure in the 
spectral variable $z$ is brought to bear.  For example, assuming the functions $b$, $b_x$, etc., to have Laurent expansions in $z$, that is to say $b=\sum b_{n}z^{n}$, $b_{x}=\sum b_{n,x}z^{n}$, and $b_{xxx}=\sum b_{n,xxx}z^{n}$, then collecting like powers of $z$ in \eqref{eq:deformationeq} gives
\begin{equation} \label{eq:compatibility}
       \rho_{t}=\frac{1}{2}b_{1,xxx}+\In b_0,\qquad \frac{1}{2}b_{n,xxx}+\In b_{n-1}=0, \qquad \text{for} \ n\neq 1. 
\end{equation} 
If one assumes $b$ to be a rational function
\begin{equation}\label{eq:rationalD}
    b(z)=b_0+\sum_{k=1}^{M}\frac{b_{-1}^{(k)}}
    {z+\epsilon_k}, \qquad \epsilon_k\neq \epsilon_j, \qquad k\neq j \text{ and } 
    \epsilon_k>0
\end{equation} 
then the deformation equation \eqref{eq:deformationeq} reads: 
\begin{subequations} 
\begin{align}
    \rho_{t}&=\In b_0, \label{eq:CGSratt}\\
    b_{0,xxx}+\sum_{k=1}^M \frac{b_{-1,xxx}^{(k)}}{\epsilon_k}=0, &\qquad 
    \frac{1}{2}b_{-1,xxx}^{(k)} -\epsilon_k \In b_{-1}^{(k)}=0, \quad k=1\dots M.  \label{eq:CGSratcons}
\end{align} 
\end{subequations}

Let us consider a few specific examples of deformation equations.  
\begin{example} \label{ex:HD} 
Taking $b = zb_{1}+b_{0}$ one obtains
\begin{equation}\label{eq:HD}
    \rho_{t}=\frac{1}{2}b_{1,xxx}+\In b_{0}, \qquad 
    \In b_{1} = 0, \qquad b_{0,xxx} = 0.  
\end{equation}
This deformation was mentioned already in \cite{Sabatier-Evolution} where this evolution equation was identified 
as the {\sl Harry-Dym equation} ~(HD).  
\end{example}
\begin{example}\label{ex:CH}
Similarly, the case $b = b_{0}+\frac{b_{-1}}{z}$ results in
\begin{equation}\label{eq:CH} 
\rho_{t}=\In b_{0}, \qquad 
	\frac{1}{2}b_{0,xxx}+\In b_{-1} = 0, \qquad 
	b_{-1,xxx} = 0.
\end{equation}
The relation of this deformation to the string with Dirichlet boundary conditions was 
established in \cite{BSS-Stieltjes} and, in this special case, 
the evolution equation \eqref{eq:CH} (after a Liouville transformation) corresponds to the Camassa-Holm ~(CH) equation \cite{CH} mentioned earlier.    
 
\end{example} 
Finally we have the rational model, the main focus of this paper.  
\begin{example} \label{ex:RM}
We put $M=1$, set $\, \epsilon_1=\epsilon, \, b_{-1}^{(1)}=b_{-1} $ in \eqref{eq:rationalD} and rearrange equations to obtain
\begin{equation} \label{eq:CGS}
    \rho_{t}=\In b_0, \qquad 
    \frac{1}{2}b_{0,xxx} +\In b_{-1}=0, \qquad \epsilon b_{0,xxx}+b_{-1,xxx}=0.  
    \end{equation} 
 \end{example}    

\begin{remark} 
The idea behind the use of the parameter $\epsilon$ is that we want to have control over the singularity of $b$, in particular 
we do not want to have a singularity at a point of the spectrum of the string. 
 Moreover, we plan to compute the limit $\epsilon \rightarrow 0$ in the 
 rational model.  The presentation of example \ref{ex:RM} suggests 
 that, at least formally, the rational model goes over to the one discussed 
 in example \ref{ex:CH}.  Finally, the relevance of the rational model, 
 in addition to the fact that it goes in the limit $\epsilon \rightarrow 0$ to 
 other known cases, is that it admits a natural resolution of the 
 differential constraints in terms of products of eigenfunctions, exemplified 
 by Theorems \ref{thm:bconstruction}, \ref{thm:bconstructionD}, or 
 in the general case of rational flows \eqref{eq:rationalD}, by Theorem \ref{thm:bratconstruction}.    
 
\end{remark} 
  
In the remainder of this section we discuss one type of non-smooth mass densities, 
exemplified by $\rho=\sum_{j=1}^N m_j\delta_{x_j}$.

Away from the support of $\rho$ we are dealing with the smooth case so 
our previous discussion applies.  However, on the support of $\rho$ we 
need to apply distributional (in the sense of theory of distributions) calculus.  
In Appendix \ref{app:weak formulation}, we analyze the deformation equation on the support of $\rho$ and 
show that the deformation equation will take the form of the distributional 
equation
\begin{equation}\label{eq:Ddeformationeq}
z D_t\rho=\frac12 D^3_x b+z\In b
\end{equation} 
where $D_t, D_x$ are distributional derivatives in $t,x$ respectively 
while $\In f =D_x(\rho f )+\rho  \langle f_x \rangle$ for any continuous piecewise smooth function $f$ with $\langle f_x \rangle$ denoting the average function (the 
arithmetic average of the right-hand and left-hand limits).   
\begin{example} 
For the rational flow $b=b_0+\frac{b_{-1}}{z+\epsilon}$ and 
$\rho=\sum_{j=1}^N m_j\delta_{x_j}$ the deformation equations 
restricted to the support of $\rho$ read: 
\begin{equation} \begin{gathered} 
D_t \rho=\sum_{j=1}^N \dot m_j \delta_{x_j} - m_j \dot x_j \delta_{x_j}', \\
zm_j\dot x_j=-\frac12 [b_x](x_j)-zm_j b(x_j),\\
z\dot m_j=\frac12 [b_{xx}](x_j)+zm_j\langle b_x \rangle (x_j), 
\end{gathered}
\end{equation}
where $[f](x_j)$ is the jump of $f$ at $x_j$.  
Writing these equations in terms of components $b_0$ and $b_{-1}$ 
we obtain: 
\begin{align} 
    \dot x_j&=-b_0(x_j),  \quad &-\epsilon [b_{0,x}](x_j)&=[b_{-1,x}](x_j), \quad  
    &\frac12 [b_{-1,x}](x_j)&=\epsilon m_ib_{-1}(x_j)\\
    \dot m_j&=m_j \langle b_{0,x} \rangle (x_j), \quad &-\epsilon [b_{0,xx}](x_j)&=[b_{-1,xx}](x_j), \quad 
    &\frac12 [b_{-1,xx}](x_j)&=\epsilon m_j\langle b_{-1,x}\rangle (x_j),   
\end{align}
producing, in the formal limit $\epsilon \rightarrow 0\, (\text {corresponding to }b=b_0+\frac{b_{-1}}{z})$, the dynamical system
similar to the one known from the CH theory: 
\begin{equation}
\dot x_j=-b_0(x_j),  \qquad \qquad \dot m_j=m_j \langle b_{0,x} \rangle (x_j).     
\end{equation}
These are only preliminary computations which will be fully justified once 
the existence of the limit for $b_{-1}$ and $b_0$ is proved.  
\end{example}
\section{Boundary conditions }\label{sec:BCs}
The boundary conditions \eqref{eq:stringBVP} can be written as 
\begin{equation*}
	 \begin{bmatrix} -h&1 \end{bmatrix} V(x=0) = 0, \qquad 
	\begin{bmatrix} H& \, \, 1\end{bmatrix} V(x=1) = 0.  
\end{equation*}
We require that these boundary conditions hold during the deformation of the string, 
in other words we require 
\begin{equation*}
\begin{bmatrix} -h& 1\end{bmatrix} V_{t}(x=0) = 0, \qquad 
	\begin{bmatrix} H &\, \,  1 \end{bmatrix} V_{t}(x=1) = 0,
\end{equation*}
or 
\begin{equation}\label{eq:deformationBC}
\begin{gathered}
	\begin{bmatrix} -h& 1\end{bmatrix} \begin{bmatrix} a&b\\c&d \end{bmatrix}V(x=0) = 0, \\
	\begin{bmatrix} H& \, \, 1 \end{bmatrix} \begin{bmatrix} a&b\\c&d \end{bmatrix}V(x=1) = 0,
\end{gathered}
\end{equation}
where $a$, $c$, and $d$ have been determined in \eqref{eq:abc} in terms of the function $b$,  thus leading to the explicit, $b$-dependent, form of \eqref{eq:deformationBC}.

\begin{lemma}\label{lem:BCinvariance}  
Suppose $v$ satisfies the string boundary value problem \eqref{eq:stringBVP}  at time 
$t$.  Then the ZS deformation leaves the boundary conditions invariant provided 
\begin{equation}\label{eq:GenBC1}
\begin{split}
	& \frac12 b_{xx}(0)-hb_x(0)+(h^2+z \rho(0))b(0)= 0, \\
	& \frac12 b_{xx}(1)+H b_x(1)+(H^2+z\rho(1)) b(1) = 0.  
\end{split}
\end{equation}
In particular, if there is no mass at the endpoints then 
\begin{equation}\label{eq:sGenBC1}
\begin{split}
	& \frac12 b_{xx}(0)-hb_x(0)+h^2 b(0)= 0, \\
	& \frac12 b_{xx}(1)+H b_x(1)+H^2 b(1) = 0, 
\end{split}
\end{equation}
with the proviso that $b(0)=0$ if $h=\infty$, likewise $b(1)=0$ if $H=\infty$.   
\end{lemma}
\begin{remark} We emphasize that condition \eqref{eq:GenBC1} has to 
hold not only in $t$ but also in $z$, resulting in conditions on the 
components of $b$.  The latter leads to severe restrictions on admissible 
choices of $b$, for example, eliminating the polynomial flows
as the example below illustrates.  
\end{remark} 
We briefly analyze a few special cases of Lemma \ref{lem:BCinvariance}.  
\begin{example} 
[Neumann and Dirichlet Conditions]
The Neumann-Neumann  boundary conditions have $h=H= 0$, hence 
$b_{xx}(0)=b_{xx}(1)=0$, in the case of no masses at the end points.  
In the Dirichlet-Dirichlet  case, regardless of the presence of masses at the endpoints, $h=H=\infty$, hence $b(0)=b(1)=0$.   
\end{example} 
We illustrate the relevance of Lemma \ref{lem:BCinvariance} in highlighting 
the difference brought about by specific analytic properties of $b$ at 
$z=\infty$. 
In particular, we would like to gauge to which extent it is possible to impose the boundary conditions ~\eqref{eq:GenBC1} on 
~$b$ without violating the differential equation ~\eqref{eq:deformationeq}.  To appreciate the task we will analyze two cases.  

\noindent {\bf Polynomial flows: $\mathbf{b = b_{1}z+b_{0}} $ }

The differential equations \eqref{eq:HD} yield solutions of the form
\begin{equation}\label{eq:polybs}
	b_{0}(x)=Ax^{2}+Bx+C, \qquad \qquad b_{1}(x) = \frac{D}{\rho(x,t)^{1/2}}. 
\end{equation}

Suppose we impose Dirichlet-Dirichlet boundary conditions.  Then 
Lemma \ref{lem:BCinvariance} implies  $b_{0}(0)=b_{0}(1)=0$ and $b_{1}(0)=b_{1}(1)=0$, hence $b_0(x)=Ax(1-x)$ while by the second condition either 
$D=0$ or  
\begin{equation*}
	b_{1}(0)=b_1(1)=\frac{D}{\rho(0,t)^{1/2}}=\frac{D}{\rho(1,t)^{1/2}}=0.
\end{equation*}	
Thus the polynomial flow does not preserve Dirichlet-Dirichlet boundary conditions unless 
the density is singular at the endpoints.  Consequently, for general 
densities, $D=0$, hence $b_1(x)=0$, and the only polynomial flow preserving 
Dirichlet-Dirichlet boundary conditions will be linear: $\rho_{t}=~\In b_{0}, \, \,  b_0=Ax(1-x)$.   
To deal with boundary conditions other than Dirichlet-Dirichlet boundary conditions we observe that 
by inspection equation \eqref{eq:GenBC1} implies 
$\frac12 b_{i, xx}(0)-hb_{i,x}(0)+h^2 b_i(0)+\rho(0) b_{i-1}(0)=0$.   
So, in the case of $i=2$, if there is a mass at the left endpoint, $b_1(0)=0$ which implies $D=0$ unless 
the density is singular at $x=0$.  Thus in the regular case again $b_1(x)=0$ and we conclude that for the general boundary conditions and general 
mass densities the linear flow above is the only 
flow preserving these boundary conditions.   
This argument can easily be generalized to the case of polynomial flows $\sum_{n=0}^N b_nz^n$.  We conclude this example by pointing out 
that because of the form of $b_1$ which involves both taking the root and inverting of $\rho$, one does not expect the situation to improve for $\rho$ given by discrete measures.  

  Next, we illustrate how the behaviour of $b$ at $z=\infty$ impacts analysis.

\noindent {\bf Flows polynomial in $1/z$: $\mathbf {b = b_0+\frac{b_{-1}}{z}}$ }

For simplicity we will only analyze the case of vanishing mass density at the endpoints.  
Recall the system of differential equations for the case $b = b_0+\frac{b_{-1}}{z}$ (equation 
\eqref{eq:CH}): 
\begin{equation*}
        \rho_{t}=\In b_{0}, \qquad 
	\frac{1}{2}b_{0,xxx}+\In b_{-1}= 0, \qquad 
	\frac{1}{2}b_{-1,xxx} = 0.
\end{equation*}
The last equation gives $b_{-1}(x)=\frac12 \alpha_1 x^{2}+\alpha_2 x+\alpha_3$.  
The boundary conditions on $b_{-1}$ given by equations \eqref{eq:sGenBC1} 
imply: 
\begin{equation}\label{eq:C0equation}
\begin{bmatrix} \frac12 & -h & h^2\\ \frac12 +H+\frac12 H^2 & H+H^2& H^2 \end{bmatrix} 
\begin{bmatrix} \alpha_1\\\alpha_2\\\alpha_3\end{bmatrix}=\begin{bmatrix} 0\\0 \end{bmatrix}.  
\end{equation}
 These equations, by virtue of representing two planes going through the origin, will always have a solution.

 Now, as to $b_{0}$ we observe that one can write $b_{0}$ in terms of $\rho$ and $b_{-1}$ by integrating
\begin{equation}\label{eq:b0equation}
\begin{split}
	& b_{0,xx}=-2\int\rho_{x}b_{-1} dx -4\int\rho b_{-1,x}dx + C_1 = -2\rho b_{-1} -2\int\rho b_{-1,x}dx +C_{1}, \\
	& b_{0,x}=-2\int\biggl(\rho b_{-1}+\int\rho b_{-1,x}dx\biggr)dx+C_{1}x+C_{2}:= f(x)+C_{1}x+C_{2}, \\
	& b_{0}=\int f(x) dx+\frac{1}{2}C_{1}x^{2}+C_{2}x+C_{3}.
\end{split}
\end{equation}
Observe that we can always choose $f(x)$ above in such a way 
that $b_0(0)=C_3, \, b_{0,x}(0)=C_2,\,  b_{0,xx}(0)=C_1$ and then set 
$b_0(x)=\int_0^x f(\xi)d\xi +\frac{1}{2}C_{1}x^{2}+C_{2}x+C_{3}$
Then the boundary conditions on $b_{0}$ given by equations \eqref{eq:sGenBC1} 
imply: 
\begin{equation}\label{eq:Cequation}
\begin{bmatrix} \frac12 & -h & h^2\\ \frac12 +H+\frac12 H^2 & H+H^2& H^2 \end{bmatrix} 
\begin{bmatrix} C_1\\C_2\\C_3\end{bmatrix}=\begin{bmatrix} 0\\ C
\end{bmatrix},   
\end{equation}
where $C=\int_0^1 \rho(x) b_{-1,x}(x) dx -Hf(1)-H^2 \int_0^1 f(x) dx$.  This system of equations 
describes two planes, one of which goes through the origin.  These planes will always intersect unless they are parallel, in which case the normal vectors are proportional.  
For this to happen 
\begin{equation*}
\begin{bmatrix} \frac12 & -h & h^2  \end{bmatrix} 
=k \begin{bmatrix} \frac12 +H+\frac12 H^2 & H+H^2& H^2\end{bmatrix},   
\end{equation*}
for some constant $k$.  Since $0\leq h,H$ we see that the only case when this occurs
is the case $h=H=0$. However, if $h=H=0$ then equation
\eqref{eq:C0equation} implies that $b_{-1}$ is a linear function in $x$.  
Moreover, in this case, for equation \eqref{eq:Cequation} to have a solution, $C$ must vanish.  
Thus $\int_0^1 \rho(x) b_{-1,x} dx=0$ and it suffices to choose  $b_{-1}$ to be a constant to satisfy  \eqref{eq:Cequation}.  The remaining cases involving either $h=\infty$ or $H=\infty$ or $h=H=\infty$ can be analyzed in the same way.  In summary, 
for all $0\leq h, H\leq \infty$ one can satisfy conditions \eqref{eq:sGenBC1}. 
Finally, in the case of Neumann-Neumann conditions ($h=H=0$) 
 one recovers the Hunter-Saxton equation \cite{HS, BSS-HS}.
We will continue this analysis in later sections, in particular 
we will give a uniform construction of $b$ valid for all boundary conditions (see 
theorem \ref{thm:gCH}).

\section{Isospectrality for general boundary conditions } \label{sec:lin}

In this section we check directly that the $b$ flows, subject to \eqref{eq:sGenBC1}, are isospectral for the boundary value problem \eqref{eq:stringBVP}. 
To simplify analysis, and also guided by the inverse scattering technique, we will make the 
following assumption.    

\noindent {\bf Assumption} 

\noindent {\sl From this point onwards we will be assuming that there is no mass  in some open neighborhoods $I_0, I_1$ of the left endpoint, right endpoint, 
respectively. }

Our starting point is again \eqref{eq:deformationeq}.  We assume in this section that 
 conditions \eqref{eq:GenBC1} hold to ensure that the ZS deformations 
leave the boundary conditions invariant by Lemma \ref{lem:BCinvariance}.  The assumption on the absence of masses around the endpoints is in force.

On $I_{0}$ 
\begin{equation}\label{eq:vI0BC}
	v(x)=\begin{cases}hx+1&\text{if }0\leq h<\infty, \\x+\frac{1}{h}&\text{if }0<h\leq\infty, \end{cases}
\end{equation}
satisfies equation \eqref{eq:stringBVP}.   
Moreover, since there is no mass on $I_0$, $b_{xxx}=0$ by equation \eqref{eq:deformationeq}.  Hence $b$ is quadratic in $x$ on $I_0$.  
Evaluating the deformation $v_{t}=(\beta-\frac{1}{2}b_{x})v+bv_{x}$ for $v$ satisfying 
equations \eqref{eq:vI0BC} and \eqref{eq:GenBC1} yields: 
\begin{equation}\label{eq:betaleft}
\beta=\begin{cases}\frac{1}{2}b_{x}(0)-h b(0) &\text{if }0\leq h<\infty, \\-\frac{1}{2}b_{x}(0)+\frac{1}{2h}b_{xx}(0)&\text{if }0<h\leq\infty. \end{cases}
\end{equation}

Now we turn to analyzing the behaviour on $I_{1}$, finding from \eqref{eq:stringBVP} that
\begin{equation}\label{eq:vI1BC}
	v(x)=A_{1}(t,z)(x-1)+B_{1}(t,z), 
\end{equation}
implying that the spectrum, determined by the roots of $v_x(1)+Hv(1)=0$ will be defined by the roots of $A_{1}(t,z)+HB_{1}(t,z)=0$ or $\frac{1}{H}A_{1}(t,z)+B_{1}(t,z)=0$ according to whether $0\leq H<\infty$ or $0<H\leq\infty$ respectively. 
As on $I_{0}$ we similarly find that on $I_{1}$
\begin{equation*}
	b=\frac{1}{2}b_{xx}(1)(x-1)^2+b_{x}(1)(x-1)+b(1); 
\end{equation*}
suppressing the dependence of coefficients on $t$ and $z$ to avoid a cluttered notation.  
Collecting like powers of $(x-1)$ in the deformation $v_{t}=(\beta-\frac{1}{2}b_{x})v+bv_{x}$ as well as using equation \eqref{eq:sGenBC1} gives
\begin{align} 
\dot{A}_{1}&=(\beta+\frac{1}{2}b_{x}(1))A_{1}-\frac{1}{2}b_{xx}(1)B_{1} =
(\beta+\frac{1}{2}b_{x}(1))A_{1}+H(b_x(1)+Hb(1)) B_{1}, \label{eq:A1dot}\\
	\dot{B}_{1}&=(\beta-\frac{1}{2}b_{x}(1))B_{1}+b(1)A_{1}=(\beta-\frac{1}{2}b_{x}(1))B_{1}-\frac{1}{H^2}(Hb_{x}(1)+\frac{1}{2}b_{xx}(1))A_{1} \label{eq:B1dot}.
\end{align}
Hence when $0\leq H<\infty$
\begin{equation*}
\dot A_1 +H\dot B_1=\biggl(\beta +\frac12 b_x(1)+Hb(1)\biggr)\biggl(A_1+HB_1\biggr), 
\end{equation*}
or when $0<H\leq\infty$
\begin{equation*}
	\frac{1}{H}\dot A_{1}+\dot B_{1}=\biggl(\beta-\frac{1}{2}b_{x}(1)-\frac{1}{2H}b_{xx}(1)\biggr)\biggl(\frac{1}{H}A_{1}+B_{1}\biggr)
\end{equation*}
 holds.   Thus under the deformation, $v_x(1)+Hv(1)=0$ will persist, hence 
 the ZS deformation is indeed isospectral.  However, 
 the extent to which the evolution of 
 the spectral data encoded in \eqref{eq:A1dot}, \eqref{eq:B1dot}  is simpler than the original evolution 
 equation on $\rho$ is far from certain, for example the coefficient $\beta$ will in general depend (nonlocally) on $\rho$ and, in principle, could even depend on the deformation time $t$. 
 The remainder of the paper is devoted to the goal of linearizing the 
 evolution on the spectral side.  Towards this end we will rephrase 
 equations \eqref{eq:A1dot} and \eqref{eq:B1dot} by differently grouping terms on the right hand side of respective equations.

\begin{proposition} \label{prop:cond-isospectrality}
Let $v$ satisfy the string boundary value problem \eqref{eq:stringBVP} for the mass 
density $\rho$ with mass gaps around the endpoints, $0\leq h, H\leq \infty$ and let 
the ZS deformation satisfy the deformation equation \eqref{eq:deformationeq} with $b$ subject to \eqref{eq:sGenBC1}.  Let us define a function $K$ by 

\begin{equation}\label{eq:Kequation}
	K\beta = \begin{cases}\frac{1}{2}b_{x}(1)+Hb(1)&\text{ if }\, 0\leq H<\infty, \\ -\frac{1}{2}b_{x}(1)-\frac{1}{2H}b_{xx}(1)&\text{ if }\, 0<H\leq\infty,\end{cases}
\end{equation}
where $\beta$ is given by \eqref{eq:betaleft}.   Then the deformation yields the following evolution equations of the spectral data $A_1, B_1$:
\begin{enumerate}
\item For $\, 0<H<\infty$:
	\begin{subequations} 
	\begin{align}
	\dot A_1 +H\dot B_1&=(K+1)\beta [A_1+HB_1], \\
	\dot A_1-H\dot B_1&=(1-K)\beta [A_1-HB_1]+b_x(1)[A_1+HB_1], 
	\end{align}
	\end{subequations} 
\item For $H=0$:
	\begin{subequations}
	\begin{align}
	\dot A_1 &=(1+K)\beta A_1, \\
	\dot B_1&=(1-K)\beta B_1+b(1)A_1, 
	\end{align}
	\end{subequations}
\item For $H=\infty$:
	\begin{subequations}
	\begin{align}
	\dot{A}_{1} &=(1-K)\beta A_{1}-\frac{1}{2}b_{xx}(1)B_{1},\\
	\dot{B}_{1} &=(1+K)\beta B_{1}.
	\end{align}
	\end{subequations}
\end{enumerate}	
\end{proposition} 
\begin{remark} 
The objective of the next section is to show that one can choose 
the field $b$, without violating \eqref{eq:sGenBC1}, in such a way that $K=-1$ and 
$\beta$ is invariant under $t$, thereby making it computable from $\rho(t=0)$.  
\end{remark}

\section{Isospectrality and linearization for rational flows and general boundary conditions } \label{sec:bplus}
As we observed in section \ref{sec:BCs} polynomial ZS flows are not suitable 
deformations of a string boundary value problem for mass densities
regular at the endpoints, and also for those for which the mass density 
has zeros (see \eqref{eq:polybs}).  On the other hand, 
the flow of the CH type, namely $b=b_0+\frac{b_{-1}}{z}$, works fine in either case.  So 
the most natural first step towards a generalization is simply to try $b=b_0+\sum_{j=1}\frac{b_{-j}}{z^j}$.  
As stated earlier,  (see \eqref{eq:compatibility}) one then obtains 
\begin{equation*}
\rho_t=\In b_0, \qquad \frac 12 b_{0,xxx}+\In b_{-1}=0, \quad \frac 12 b_{-n, xxx}+\In b_{-(n+1)}=0, \qquad 1\leq n. 
\end{equation*}
The truncation at $n=2$, which amounts to setting 
$b_{-n}=0$ starting with $n=2$, produces the CH type system, with two interacting components 
$b_0$ and $b_{-1}$.  There is however another scenario in which one solves 
the constraints $\frac 12 b_{-n, xxx}+\In b_{-(n+1)}=0, \, 1\leq n$, not by truncation 
but by a sort of ``quasi-periodicity", positing $b_{-(n+1)}=wb_{-n}, \, 2\leq n$, 
for some proportionality constant $w$ (taken to be $-\epsilon$ in the remainder of this paper).    
Then out of the infinite set of equations one needs only 
two, namely, 
\begin{equation*}
\frac12 b_{0,xxx}+\In b_{-1}=0, \quad \frac 12 b_{-1, xxx}+w \In b_{-1}=0, 
\end{equation*}
and the theory remains described by two fields only.  This is one of the motivations 
for studying the rational model.

We will argue now that the rational flows are free of the deficiencies present for polynomials flows described 
above; in particular the vanishing of $\rho$ is no longer an issue. To this end we will  analyze the simplest rational flow with one pole.  
Thus $b=b_0+\frac{b_{-1}} {z +\epsilon}, \, \epsilon>0$ (equation \eqref{eq:rationalD}), 
 For convenience we recall the differential equations \eqref{eq:CGS} in the special case of one pole, writing them in a form convenient for further analysis: 

\begin{subequations} \label{eq:CGS1}
\begin{align}
    \rho_{t}&=\In b_0,  \label{eq:CGS1t}\\
    b_{0,xxx}+\frac{b_{-1,xxx}}{\epsilon}=0, &\qquad 
    \frac{1}{2}b_{-1,xxx}^{} -\epsilon\In b_{-1}=0 \label{eq:CGS1cons}.  
    \end{align}
\end{subequations} 
  
We begin our analysis by formulating three auxiliary lemma, the first of which can be found for example in \cite{Appell}, and is discussed in \cite{Barrett},\cite{Tefteller}.

\begin{lemma}\label{lem:solspace}
Let $\lambda \in \mathbb{C} , \, \rho \in C^1([0,1])$ and suppose $\phi$ and $\psi$ are solutions of $f_{xx}-\lambda \rho f=0$. Then $\phi^2$, $\phi\psi$, ~and $\psi^2$  
are solutions of $\frac12g_{xxx}-\lambda \In g= 0$. If, in addition, $\phi$ and $\psi$ are 
linearly independent, then $\phi^2$, $\phi\psi$, and $\psi^2$ span the solution space of $\frac12g_{xxx}-\lambda \In g= 0$. 
\end{lemma}
\begin{proof}
	By direct computation we have
	\begin{equation*}
	\begin{split}
		& (\phi\psi)_{xxx}-\lambda(2\rho_{x}\phi\psi+4\rho(\phi\psi)_{x}) \\
		=\,& \phi_{xxx}\psi+3\phi_{xx}\psi_{x}+3\phi_{x}\psi_{xx}+\phi\psi_{xxx}-\lambda(4\rho\phi_{x}\psi+4\rho\phi\psi_{x}+2\rho_{x}\phi\psi) \\
		=\,&(\phi_{xx}-\lambda\rho\phi)_{x}\psi+(\psi_{xx}-\lambda\rho\psi)_{x}\phi+3(\phi_{xx}-\lambda \rho\phi)\psi_x+3(\psi_{xx}-\lambda \rho\psi)\phi_x=0,
	\end{split}
	\end{equation*}
	and also 
	\begin{equation*}
		W(\phi^2,\phi\psi,\psi^2)=2W(\phi,\psi)^3,
	\end{equation*}
	where $W$ denotes the Wronskian. The first computation shows that these indeed form solutions of the third order problem, while the second computation shows they span its solution space provided $\phi, \psi$ are linearly independent.
\end{proof}

\begin{lemma}\label{lem:b1solgen}
	Let $\lambda \in \mathbb{C}, \, \rho\in C^1([0,1])$ and suppose $\rho(0)=\rho(1)=0$. 
	Moreover,  let $\bb=\phi\psi$ where $\phi$ and $\psi$ satisfy
	\begin{equation*}
	\begin{split}	
		&\phi_{xx}=\lambda \rho\phi,\qquad \phi_{x}(0)-h\phi(0)=0,\\
		&\psi_{xx}=\lambda \rho\psi,\qquad \psi_{x}(1)+H\psi(1)=0.
	\end{split}
	\end{equation*}

	Then $\bb$ satisfies 
	\begin{equation}\label{eq:SelfAdjointThird}
		\frac12 \bb_{xxx}-\lambda\In \bb=0,
	\end{equation}
	with the boundary conditions \eqref{eq:sGenBC1}\rm{:}
	\begin{equation*}
	\begin{split}
		&\frac{1}{2}\bb_{xx}(0)-h\bb_{x}(0)+h^2\bb(0)=0,\\
		&\frac{1}{2}\bb_{xx}(1)+H\bb_{x}(1)+H^2\bb(1)=0.
	\end{split}
	\end{equation*}
\end{lemma}
\begin{proof}
	That such a $\bb$ satisfies the third order differential equation is guaranteed by Lemma \ref{lem:solspace}. To check the boundary conditions we calculate them explicitly. In particular,  $\bb_{xx}=2\phi_{x}\psi_{x}$ at $x=0,1$ because $\rho=0$ at $x=0,1$. Then at $x=0$ \eqref{eq:sGenBC1} becomes
\begin{equation*}
	(\phi_{x}\psi_{x}-h\phi_{x}\psi-h\phi\psi_{x}+h^{2}\phi\psi)\rvert_{x=0}=(\psi_x-h\psi)(\phi_{x}-h\phi)\rvert_{x=0}=0,
\end{equation*}
and similarly at $x=1$ 
\begin{equation*}
	(\phi_{x}\psi_{x}+H\phi_{x}\psi+H\phi\psi_{x}+H^{2}\phi\psi)\rvert_{x=1}=(\phi_x+H\phi)(\psi_{x}+H\phi)\rvert_{x=1}=0.
\end{equation*} 
\end{proof}

Our attention will now turn to the determination of $\beta$ (see \eqref{eq:betaleft}) and, subsequently, $K$ from Proposition \ref{prop:cond-isospectrality}.   

First we recall the definition of the \emph{bilinear concomitant} (\cite{Ince}, p.125) associated with the third order differential equation $y'''=0$.  
\begin{definition}
Let $f,g \in C^3$ then the bilinear concomitant  is 
\begin{equation*}
B(f,g)(x)=f_{xx}g-f_xg_x+fg_{xx}.  
\end{equation*}
\end{definition}  

The second pertinent fact is the skew-symmetry of the operator $\In$ relative
to the natural inner product $(f,g)=\int_0^1fgd\, x$.  Indeed, whenever 
$\rho(0)=\rho(1)=0$ as one easily checks using integration by parts:
\begin{equation*}
(f,\In g)=-(\In f, g)
\end{equation*}
holds.

To proceed further we establish two important lemmas.

\begin{lemma}\label{lem:b1xGen}
Let $\lambda \in \mathbb{C}, \, \rho\in C^1([0,1])$ and suppose $\rho(0)=\rho(1)=0$. 
	If $\bb$ satisfies $\frac12 \bb_{xxx}-\lambda \In \bb=0$ with boundary conditions \eqref{eq:sGenBC1} then 
	\begin{equation}\label{eq:SqrEqual}
	\begin{split}
		 \biggl(\frac{1}{2}\bb_{x}(0)-h\bb(0)\biggr)^2 &=\biggl(\frac{1}{2}\bb_{x}(0)-\frac{1}{2h}\bb_{xx}(0)\biggr)^2\\
		&= \biggl(\frac{1}{2}\bb_{x}(1)+H\bb(1)\biggr)^2=\biggl(\frac{1}{2}\bb_{x}(1)+\frac{1}{2H}\bb_{xx}(1)\biggr)^2,
	\end{split}
	\end{equation}
	where the first and the third equalities are displayed in order to deal with the cases of $h=\infty,\,  H=\infty$ respectively.  
	\end{lemma}
\begin{proof}
	Since $\bb$ satisfies $\frac12 \bb_{xxx}=\lambda\In \bb$
	we easily obtain, using the skew-symmetry of $\In$, that $0=(\bb,\bb_{xxx})=~\frac12B(\bb,\bb)|_0^1$  which is equivalent to 
	\begin{equation}\label{eq:SrqEqalProof1}
		\frac{1}{2}\bb_{x}^2(1)-\bb(1)\bb_{xx}(1)=\frac{1}{2}\bb_{x}^2(0)-\bb(0)\bb_{xx}(0).  
	\end{equation}
	 Next we solve \eqref{eq:sGenBC1} for $\bb(j)$ or $\bb_{xx}(j)$, with  $j=0,1$, depending on whether we want to consider finite or infinite $h$ or $H$, subsequently obtaining 
	\begin{equation*}
	\bb(0)=\frac{1}{h}\bb_{x}(0)-\frac{1}{2h^2}\bb_{xx}(0),\qquad 
	\bb_{xx}(0) = 2h\bb_{x}(0)-2h^2\bb(0),
	\end{equation*}
and likewise at $x=1$ with $h$ replaced by $-H$. The result follows upon substituting these expressions back into \eqref{eq:SrqEqalProof1} and simplifying.
\end{proof}

\begin{lemma} \label{lem:bnonzero}
	Let $\bb$ be given as in Lemma \ref{lem:b1solgen}
	and suppose $\lambda> 0$. Then
	\begin{enumerate} 
	\item[\rm{(1)}] $\frac{1}{2}\bb_{x}(0)-h\bb(0)\neq 0, \quad $ for $\, 0\leq h<\infty$ \,  or \,   $ \frac{1}{2}\bb_{x}(0)-\frac{1}{2h}\bb_{xx}(0)\neq 0,$ \quad for $\,  0<h\leq \infty$;
	\item[\rm{(2)}] $\frac{1}{2}\bb_{x}(1)+H\bb(1)\neq 0, \quad $ for $ \, 0\leq H<\infty $\, or \,  $ \frac{1}{2}\bb_{x}(1)+\frac{1}{2H}\bb_{xx}(1)\neq 0, \quad $ for $\,  0<H\leq \infty$.  
	\end{enumerate} 
	If $\lambda=0$ inequalities (1) and (2) hold with the exception of Neumann-Neumann boundary conditions $H=h=0$ in which case $\bb_x(0)=\bb_x(1)=0$.  
	\end{lemma}
\begin{proof}

	First, let us consider the second statement of item (1).  Since by Lemma \ref{lem:b1solgen} $\bb=\phi\psi$ we can write
	\begin{equation} \label{eq:betah}
	\begin{split}
		\frac{1}{2}\bb_{x}(0)-\frac{1}{2h}\bb_{xx}(0)&=\left(\frac{1}{2}\phi_{x}\psi+\frac{1}{2}\phi\psi_{x}-\frac{1}{h}\phi_{x}\psi_{x}\right)\rvert_{x=0}\\
	&= \left(\frac{\phi_{x}}{2}\biggl(\psi-\frac{1}{h}\psi_{x}\biggr)+\frac{\psi_{x}}{2}\biggl(\phi-\frac{1}{h}\phi_{x}\biggr)\right) \rvert_{x=0}=\left(\frac{\phi_{x}}{2}\biggl(\psi-\frac{1}{h}\psi_{x}\biggr)\right) \rvert_{x=0}.
	\end{split}
	\end{equation}
	Thus it remains to show that both $\phi_{x}(0)\neq 0$ and $\psi(0)-\frac{1}{h}\psi_{x}(0)\neq0$. Indeed, $\phi_{x}(0)\neq 0$ since $\phi(0)-\frac{1}{h}\phi_{x}(0)=0$, and $h>0$ means that if $\phi_{x}(0)=0$ then also $\phi(0)=0$ which would yield only trivial solutions. Now to show $\psi(0)-\frac{1}{h}\psi_{x}(0)\neq0$  we multiply the corresponding second order problem by $\psi$ and integrate from $0$ to $1$
	\begin{equation*}
		\int_{0}^{1}\psi\psi_{xx}dx=\lambda\int_{0}^{1}\rho\psi^{2} dx \Longleftrightarrow \psi\psi_{x}\rvert_{0}^{1}-\int_{0}^{1}\psi_{x}^{2}dx=\lambda\int_{0}^{1}\rho\psi^{2}dx.
	\end{equation*}
	Solving for $\lambda$ and substituting in the boundary condition on the right gives
	\begin{equation*}
		\lambda = -\frac{H\psi(1)^2+\psi(0)\psi_{x}(0)+\int_{0}^{1}\psi_{x}^2dx}{\int_{0}^{1}\rho\psi^{2}dx},
	\end{equation*}
	from which we see that if $\psi(0)-\frac{1}{h}\psi_{x}(0)=0$ then 
	\begin{equation*}
		\lambda = -\frac{H\psi(1)^2+\frac{1}{h}\psi_{x}(0)^{2}+\int_{0}^{1}\psi_{x}^2dx}{\int_{0}^{1}\rho\psi^{2}dx}\leq 0, 
	\end{equation*}
	contradicting our  our assumption $0<\lambda$. 
	
	For the first statement of item (1), we begin by computing
	\begin{equation}\label{eq:betaphi0}
		\frac{1}{2}\bb_{x}(0)-h\bb(0)=\frac{\phi(0)}{2}\biggl(\psi_{x}(0)-h\psi(0)\biggr),
	\end{equation}
	and then proceed in a similar way to the argument in support of the second statement of item (1).  
	
	For item (2) we find that
	\begin{equation}\label{eq:betaphi1}
	\begin{split}
		&\frac{1}{2}\bb_{x}(1)+\frac{1}{2H}\bb_{xx}(1) = \frac{\psi_{x}(1)}{2}\biggl(\phi(1)+\frac{1}{H}\phi_{x}(1)\biggr),\\
		&\frac{1}{2}\bb_{x}(1)+H\bb(1)=\frac{\psi(1)}{2}\biggl(\phi_{x}(1)+H\phi(1)\biggr),
	\end{split}
	\end{equation}
	from which the same conclusions as in the last two cases can be made by similar arguments. Finally the case $\lambda=0$ can be checked by 
	direct computation using $\phi(x)=hx+1, \psi(x)=H(1-x)+1$.  
\end{proof}

\begin{lemma}\label{lem:sb1xGenRevised}
	Let $\bb=\bb(x;\lambda)$ be given as in Lemma \ref{lem:b1solgen}.  
	 Then
	\begin{enumerate}
	\item $\bb$ is entire in $\lambda$ of order $\frac12$ ;
		\item the following equalities hold (whenever they make sense) 
		\begin{equation}\label{eq:sb1xEq2}
		\begin{split}
			\biggl(\frac{1}{2}\bb_{x}(0)-\frac{1}{2h}\bb_{xx}(0)\biggr)=&-\biggl(\frac{1}{2}\bb_{x}(0)-h\bb(0)\biggr)\\
		=&-\biggl(\frac{1}{2}\bb_{x}(1)+\frac{1}{2H}\bb_{xx}(1)\biggr)=\biggl(\frac{1}{2}\bb_{x}(1)+H\bb(1)\biggr).
		\end{split}
		\end{equation}
	\end{enumerate}
	\end{lemma}

\begin{proof}
\begin{enumerate}
	\item[(1)] By Lemma \ref{lem:b1solgen} it suffices to show that the solution to $\phi_{xx}-\lambda\rho\phi=0$ with $\phi_{x}(0)-h\phi(0)=0$ or $\phi_{x}(1)+H\phi(1)=0$ is analytic in $\lambda$. This part is well known but we will need 
	some intermediate formulas so, for completeness, we proceed with the proof,
	referring to the monograph \cite{PT} for details.   
	  We begin by writing $\phi$ as a power series in $\lambda$: 
	\begin{equation*}
		\phi(x;\lambda)=c_{0}(x)+\sum_{n=1}^{\infty}\lambda^{n}c_{n}(x),
	\end{equation*}
which when substituted into the second order differential equation yields the system
\begin{equation*}
	c_{0,xx}=0,\qquad c_{n,xx}-c_{n-1}\rho=0\;\;(n\geq1).
\end{equation*}
Since the argument proceeds in the same way regardless of whether we use the 
left or right initial condition we concentrate now on the left initial condition.  

We immediately see that $c_{0}$ will be linear in $x$ with two unknowns which can be chosen to satisfy the initial condition on the left. Then we choose $c_{n}$ so that $c_n(0)= c_{n,x}(0)=0, $ holds for $ n \geq1$, ensuring that $\phi$ satisfies the left initial condition. Once this is accomplished we must show that the power series converges.

To determine $c_{n}$ ($n\geq1$) we simply integrate the $c_{n}$ relation from $0$ to $x$. Then by recursion we find that 
\begin{equation} \label{eq:cn}
c_n(x)=\int_{0\leq \xi_1\leq \xi_2\leq \dots \leq \xi_{n+1}=x}\left [\prod_{j=1}^n(\xi_{j+1}-\xi_j)\rho(\xi_j)\right] c_0(\xi_1)d\xi_1d\xi_2\dots d\xi_n
\end{equation} 

To show that the power series converges it suffices to observe that
\begin{equation}\label{eq:sb1xGenEq1}
\begin{split}
	|c_{n}(x)| & \leq Cx^n\int_{0\leq \xi_1\leq \xi_2\leq \dots \leq \xi_{n+1}=x}\left [\prod_{j=1}^n\rho(\xi_j)\right]d\xi_1d\xi_2\dots d\xi_n \\
	& \leq C \frac{x^n}{n!}\left(\int_{[0,x]}\rho(\xi)d\xi\right)^n\stackrel{\emph{def}}{=}C \frac{x^n}{n!}(M(x))^n, 
\end{split}
\end{equation}
where $|c_0(x)|<C$ and $M(x)$ is the total mass on the interval $[0,x]$.  Hence,  
the power series $\phi(x;\lambda)$ converges uniformly on compact sets in $\lambda\in\mathbb{C}$. One can improve on \eqref{eq:sb1xGenEq1}
by estimating more accurately the term $(x-\xi_n)\prod_{j=1}^{n-1}(\xi_{j+1}-\xi_j)$
in the region $\{0\leq \xi_1\leq \xi_2\leq \dots \xi_n \leq x\} $. For example, 
one can check by induction that 
\begin{equation*} 
(x-\xi_n)\prod_{j=1}^{n-1}(\xi_{j+1}-\xi_j)\leq \left(\frac{x}{n}\right)^n
\end{equation*} 
and, upon replacing $x^n$ in \eqref{eq:sb1xGenEq1} with this more accurate estimate, we obtain 
\begin{equation*}
|c_n|(x)\leq C \frac{(xM(x))^n}{n!n^n}.  
\end{equation*}  
Moreover, since $2^n n^n n!> (2n)!$, one easily obtains 
\begin{equation*}
   |\phi(x;\lambda)|\leq C \cosh \left(\sqrt{2xM(x)|\lambda| }\right), 
\end{equation*} 
implying that $\phi(x;\lambda)$ is an entire function of order not exceeding $\frac12$.  That the order is exactly $\frac12$ requires more subtle analysis, 
for example of the growth of zeros of $\phi$, which can be found in 
M.G. Krein's work \cite{Kreinstring} and citations therein.  

Since the computation for $\psi(x;\lambda)$ is similar we record only the main steps: 
\begin{equation}\label{eq:hatcn}
\begin{gathered}
    \psi(x;\lambda)=\hat c_0(x)+\sum_{n=1}\lambda^n\hat c_n(x)\\
    \hat c_{0,xx}=0, \qquad \hat c_{n, xx}=\rho \hat c_{n-1}, \qquad \hat c_{0,x}(1)+H\hat c_0(1)=0, \quad \hat c_{n,x}(1)=\hat c_n(1)=0\\
    \hat c_n=\int_{0\leq \xi_{n+1}=x\leq \xi_n\leq \dots \leq \xi_1\leq 1}\left [\prod_{j=1}^n(\xi_{j}-\xi_{j+1})\rho(\xi_j)\right] \hat c_0(\xi_1)d\xi_1d\xi_2\dots d\xi_n. 
\end{gathered}
\end{equation} 
Finally the growth estimate reads: 
\begin{equation*}
|\hat c_n|(x)\leq \hat C \frac{((1-x)\hat M(x))^n}{n!n^n}, 
\end{equation*}
where $\hat C, \hat M(x)$ denote the bound on $\hat c_0$, the mass of $[x,1]$ 
respectively.    Thus $\psi$ is also entire of order not exceeding $\frac12$, hence so is 
the product of $\phi$ and $\psi$.  

	\item[(2)] Except for the special case $h=H=0$ (i.e. Neumann-Neumann boundary conditions) (see Lemma \ref{lem:bnonzero}) we proceed by showing first that the equalities hold when $\lambda=0$  and extend them by continuity to a neighbourhood of $\lambda=0$.  
	Then the  result for $\lambda >0$ follows by using Lemmas \ref{lem:b1xGen}, \ref{lem:bnonzero} and analyticity of $b$ as a function of $\lambda$ proved in item (1). 

Indeed,  for $\lambda=0$ 	and we can take $c_0=hx+1, \hat c_0=H(1-x)+1, \, \bb=c_0\hat c_0$, and use equations \eqref{eq:betaphi0}, \eqref{eq:betaphi1} to verify the equalities.  For boundary conditions other than Neumann-Neumann 
all quantities in brackets involving $b$ are nonzero so we can form ratios.  By Lemma \ref{lem:b1xGen} all these ratios squared are $1$ for any $\lambda$.  
Since these ratios are $-1$ for $\lambda=0$ they have to stay $-1$ in a neighbourhood of $\lambda=0$ by continuity.  Thus equations \eqref{eq:sb1xEq2}
hold in the whole domain of analyticity ($\C$) by the Identity Theorem.  

For the special case $h=H=0$ it suffices to check that the equalities hold 
in a neighbourhood of $\lambda=0$ by examining $\bb$ up to the first order in $\lambda$.  Since in this case $c_0=\hat c_0=1$ we obtain 
\begin{equation*} 
\begin{gathered} 
\bb=1+\left(c_1+\hat c_1\right)\lambda +\Oh(\lambda^2)=1+\left(\int_0^x(x-\xi)\rho(\xi) d\xi +\int_x^1(\xi-x)\rho(\xi) d\xi\right )\lambda +\Oh(\lambda^2), \\
\bb_x=\left(\int_0^x\rho(\xi)d\xi -\int_x^1 \rho(\xi)d\xi\right)\lambda +\Oh(\lambda^2),
\end{gathered}
\end{equation*}	
which, when evaluated, give $\bb_x(0)=-(\int_0^1\rho(\xi) d\xi) \,  \lambda+\Oh(\lambda^2), \, 
\bb_x(1)=(\int_0^1\rho(\xi)d\xi ) \, \lambda+\Oh(\lambda^2)$, hence proving \eqref{eq:sb1xEq2} 
for $h=0=H$.  

\end{enumerate}
\end{proof}
Recall that the eigenvalues of the original boundary value problem \eqref{eq:stringBVP} are roots of $v_x(1;z)+Hv(1;z)~\stackrel{\emph{def}}{=}D(z)$ which  
we can write either in terms of $\phi$ or $\psi$, after 
adjusting the overall multiplier, for example by agreeing that $v(x;z)=hx+1=c_0$ 
on the massless portion to the right of $0$.   Taking into account Lemma \ref{lem:solspace} 
we easily obtain: 
\begin{center} 
\boxed{ D(-\lambda)= \phi_x(1;\lambda)+H\phi(1;\lambda)=-\left(\psi_x(0;\lambda)-h\psi(0;\lambda)\right).}
\end{center} 
One useful way of computing $D$ is by using the expansion in terms of iterates $c_n$ (see equation \eqref{eq:cn}.  
\begin{lemma} \label{lem:Dproduct}
Let $\rho \in C^1([0,1])$ and let $W(f,g)=fg'-f'g$ denote the 
Wronskian of two functions $f,g$.  Suppose  that $h+H>0$ and let the eigenvalues of the string problem \eqref{eq:stringBVP} be denoted by $0<z_1<z_2<\dots$, 
then $D$ admits the additive representation
\begin{equation}\label{eq:Dformula}
D(-\lambda)=W(\hat c_0,c_0)+\sum_{n=1} \lambda^n \int_{0\leq \xi_1\leq \xi_2\cdots \xi_n\leq 1}
\hat c_0(\xi_n)\rho(\xi_n)\left(\prod_{j=1}^{n-1}(\xi_{j+1}-\xi_j) \rho(\xi_j)\right)c_0(\xi_1) d\xi_1d\xi_2\dots d\xi_n, 
\end{equation}
and the product representation
\begin{equation}\label{eq:Dformulaprod}
D(-\lambda)=W(\hat c_0,c_0)\prod_{j=1}^{\infty} \left(1+\frac{\lambda}{z_j}\right).  
\end{equation}
\end{lemma}
\begin{proof}

The additive formula follows from the definitions \eqref{eq:cn}, \eqref{eq:hatcn} of $c_n$, 
 $\hat c_n$, respectively.   
The multiplicative formula is well known and it follows immediately from 
two facts: $\phi$ is an entire function of order $\frac12$ and $\frac{z_n}{n^2}=\Oh(1), 
\, n\rightarrow \infty$, 
and it therefore admits Hadamard's product formula. 
\end{proof} 
Now we are ready to give a complete construction of $b_{-1}$ and $b_0$ 
appearing in equation \eqref{eq:CGS1}.  The notation in the Theorem below 
highlights the dependence on the parameter $\epsilon$; for more details 
regarding this notation we refer to the proof of Lemma \ref{lem:sb1xGenRevised}.  

\begin{theorem} \label{thm:bconstruction}
Let $\rho \in C^1([0,1])$, $\beta, K$ be defined as in Proposition \ref{prop:cond-isospectrality}, using equations \eqref{eq:betaleft}, \eqref{eq:Kequation}  respectively, and let $\omega$ 
be given as in Lemma \ref{lem:b1solgen}.  
Let 
\begin{equation*}
\begin{gathered}
b_{-1}(x;\epsilon)=\bb(x;\epsilon),\\
b_0(x;\epsilon)=\frac{\big[\bb(x;0)-\bb(x;\epsilon)\big]}{\epsilon},\\ 
\end{gathered} 
\end{equation*}

then $b_{-1}$ and $b_0$ so defined satisfy \eqref{eq:CGS1cons}.

Moreover, for this choice of $b_{-1}, b_0$,  
\begin{enumerate} 
\item $\beta(z;\epsilon)=\frac{1}{2} [\frac{\left( D(-\epsilon)-D(0)\right)}{\epsilon}-
\frac{D(-\epsilon)}{z+\epsilon}]$, 

\item $K=-1$, 
\item $\beta(z;\epsilon)$ is deformation invariant ($\dot \beta=0$).
\end{enumerate}    
\end{theorem} 
\begin{proof} The formula for $b_{-1}$ follows from Lemma \ref{lem:b1solgen} 
for $\lambda=\epsilon$ to conform with equations \eqref{eq:CGS1}.  Likewise, the 
formula for $b_0$ is a solution of the second equation in \eqref{eq:CGS1}.  The additional 
term involving $c_0\hat c_0$ is added to ensure the existence of the limit 
$\epsilon \rightarrow 0$.  To prove the formula for $\beta$ one needs to 
consider all individual cases.  
\begin{enumerate} 
\item $0\leq h<\infty$: we use the formula \eqref{eq:betaphi0} which we 
recall for the sake of clarity, explicitly displaying the $\epsilon$ dependence
$\frac12 b_{-1,x}(0;\epsilon)-hb_{-1}(0;\epsilon)=\frac12\phi(0;\epsilon)\left(\psi_x(0;\epsilon)-h\psi(0;\epsilon)\right)=-\frac12 D(-\epsilon)$ 
where we used that $\phi(0;\epsilon)=1$ by construction.  For 
$b_0$, using linearity, we obtain $\frac12 b_{0,x}(0;\epsilon)-hb_{0}(0;\epsilon)=
\frac{1}{2\epsilon} \left(D(-\epsilon)-D(0)\right)$
\item $0< h\leq \infty$: to compute $\beta$ in this case we use the 
second formula in equation \ref{eq:betaleft} as well as \eqref{eq:betah} to obtain
$-\frac12 b_{-1,x}(0;\epsilon)+\frac{1}{2h}b_{-1,xx}(0;\epsilon) =-\frac12 D(-\epsilon)$,
recalling that for this range of $h, c_0=x+\frac{1}{h}$; similar argument to the one used in the first item works for $b_0$.  
\end{enumerate} 
The computation of $\beta$ follows now from the definition \eqref{eq:betaleft}.  
The statement that $\beta$ is constant under the deformation, in other words 
$\beta$ is a constant of "motion", is an automatic consequence 
of isospectrality which implies that $\dot D(\lambda)=0$; in principle there could be 
an overall time-dependent factor but by construction the coefficient of $\lambda^0$ 
in $D(\lambda)$ is $(c_{0,x}+Hc_0)(1)$ which is constant in $t$.  Finally, $K=-1$  (see \eqref{eq:Kequation}~) follows from \eqref{eq:sb1xEq2}.  
\end{proof} 
We conclude this section with an appropriate modification of 
Theorem \ref{thm:bconstruction} for the rational flows \eqref{eq:rationalD}. 
The proof is a straightforward generalization of the previous proof, 
taking into account that the fields $b_0$ and $b_k$ in equation 
\eqref{eq:CGSratcons} can be rescaled.  

\begin{theorem} \label{thm:bratconstruction}
Let $\rho \in C^1([0,1])$, $\beta, K$ be defined as in Proposition \ref{prop:cond-isospectrality} using equations \eqref{eq:betaleft}, \eqref{eq:Kequation}  respectively, and let $\omega$ 
be given as in Lemma \ref{lem:b1solgen}.  
Given $M$ distinct positive numbers $\epsilon _k, 1\leq k\leq M, $
and $M+1$ functions  of $\mathbf{\epsilon}=(\epsilon_1, \dots, \epsilon_M)$, denoted $\mu_0, \mu_1, \dots, \mu_M$,  let 
\begin{equation*}
\begin{gathered}
b_{-1}^{(k)}(x)=\mu_k\bb(x;\epsilon_k), \quad 1\leq k\leq M\\
b_0(x)=\sum_{k=1}^M\frac{\big[\mu_0\bb(x;0)-\mu_k\bb(x;\epsilon_k)\big]}{\epsilon_k}.\\ 
\end{gathered} 
\end{equation*}

Then $b_{-1}^{(k)}$ and $b_0$ so defined satisfy   
\begin{equation*}
b_{0,xxx}+\sum_{k=1}^M \frac{b_{-1,xxx}^{(k)}}{\epsilon_k}=0, \qquad 
    \frac{1}{2}b_{-1,xxx}^{(k)} -\epsilon_k \In b_{-1}^{(k)}=0, \quad k=1\dots M, 
\end{equation*} 
(see \eqref{eq:CGSratcons}).
Moreover, for this choice of $b_{-1}^{(k)}, b_0$,  setting 
$b=b_0+\sum_{k=1}^M \frac{b_{-1}^{(k)}}{z+\epsilon_k}$, one obtains
\begin{enumerate} 
\item 
\begin{equation}\label{eq:betarat}
\beta(z;\epsilon)=\frac12 \sum_{k=1}^M\big[\frac{\left( \mu_kD(-\epsilon_k)-\mu_0D(0)\right)}{\epsilon_k}-
\mu_k\frac{D(-\epsilon_k)}{z+\epsilon_k}\big], 
\end{equation} 

\item $K=-1$, 
\item $\beta(z; \epsilon)$ is deformation invariant.
\end{enumerate}    
\end{theorem} 

\subsection{Discrete mass density} 
In this short section we will discuss  necessary  modifications of the formalism
for the non-smooth case of 
\begin{equation} \label{eq:drho}
    \rho=\sum_{j=1}^N m_j \delta_{x_j}, \qquad 0<x_1<\dots<x_N<1, 
\end{equation}
with weights (point masses) $m_j>0$.   We will leave the details of a general non-smooth case to future work; our goal is to argue that the formalism is valid for the case of discrete positive measures with only minor 
modifications.  We start with the discussion of the boundary value problem 
\eqref{eq:stringBVP}. The only modification needed to deal with discrete measures is to take derivatives as distributional derivatives.  Thus the boundary 
value problem simply reads: 
\begin{equation}
-D^2_x v=z\rho(x) v, \quad 0<x<1, \qquad v_x(0)-hv(0)=0, \quad v_x(1)+Hv(1)=0, 
\end{equation}
implying that in the case of the discrete $\rho$, $v$ will be 
continuous and piecewise differentiable, with jumps on the support of $\rho$.  For such functions 
the distributional derivatives can be conveniently computed from 
\begin{equation}\label{eq:D}
D_x f=f_x+\sum_{j=1}^N [f](x_j)\delta_{x_j}, 
\end{equation}
and its higher order extensions; 
in the formula \eqref{eq:D} $f_x$ denotes the classical derivative, 
$[f](x_j)$ means the jump of $f$ at $x_j$.  Thus, away from the support of $\rho$, the equation reads: 
$-v_{xx}=0$, 
while on the support of $\rho$ we obtain: 
\begin{equation*}
-[v_x](x_j)=zm_j v(x_j).  
\end{equation*}

We will also need to extend the action 
of $\In$ to continuous, piecewise differentiable functions $f$: 
\begin{equation} \label{eq:weakL}
\In f\stackrel{\emph{def}}{=} D_x(f\rho)+\langle f_x\rangle \rho=D_x(f\rho)+\sum_{j=1}^N\langle f_x\rangle (x_j)m_j  
\delta_{x_j}
\end{equation} 
where $\langle f\rangle$ is an everywhere defined average function 
of pointwise right-hand and left-hand limits.  

We will now examine appropriate modifications of Lemmas \ref{lem:b1solgen}, \ref{lem:solspace}
which are critical for the whole formalism.

 \begin{lemma}\label{lem:solspaceD}
Let $\lambda \in \mathbb{C}, \rho$ be a discrete measure given by \eqref{eq:drho} and suppose $\phi$ and $\psi$  are solutions of the distributional equation $D^2_x f-\lambda \rho f=0$. Then $\phi\psi$ is a solution of the distributional equation $\frac12D_x^3 g-\lambda \In g= 0$.
\end{lemma}
\begin{proof}
We observe that both $\phi$ and $\psi$ are continuous and piecewise differentiable; in fact piecewise linear.  Away from the support of $\rho$ the claim holds by 
the original Lemma \ref{lem:solspace}.  On the support, that is at the points $x_j$, 
one needs to compute the contributions of distributional derivatives.  
One can localize the problem by choosing test functions to be supported only 
in the neighbourhood of individual points of the support.  Hence, we can perform 
computations as if $\rho$ were $\rho=m_1\delta_{x_1}$, thereby dropping 
the sum from the computation, only to reinstate it in the final answer.  We compute 
now $D_x^3(\phi\psi)$: 
\begin{equation*}
\begin{gathered} 
D_x^3(\phi\psi)=D_x^2((\phi\psi)_x)=D_x\left((\phi\psi)_{xx}+[(\phi\psi)_x](x_1)\delta_{x_1}\right)=\\
(\phi\psi)_{xxx}+[(\phi\psi)_x](x_1)\delta'_{x_1}+[(\phi\psi)_{xx}](x_1)\delta_{x_1}=[(\phi\psi)_x](x_1)\delta'_{x_1}+[(\phi\psi)_{xx}](x_1)\delta_{x_1}, 
\end{gathered}
\end{equation*}
where we used that $\phi, \psi$ are piecewise linear (in $x$) continuous functions.
We recall that $[\phi_x](x_1)=\lambda m_1\phi(x_1), \, [ \psi_x](x_1)=\lambda m_1 \psi(x_1)$, which implies
\begin{equation*}
[(\phi\psi)_x](x_1)\delta_{x_1}'=2\lambda \phi(x_1)\psi(x_1)m_1\delta'_{x_1}=
2\lambda D_x((\phi\psi)\rho).  
\end{equation*}
Likewise, for the second singular term we obtain: 
\begin{equation*} 
\begin{split}
[(\phi\psi)_{xx}](x_1)\delta_{x_1}=2[\phi_x \psi_x](x_1)\delta_{x_1}=
2\left([\phi_x](x_1)\langle \psi_x\rangle (x_1)+[\psi_x](x_1)\langle \phi_x\rangle (x_1)\right)\delta_{x_1}=\\
2\left(\lambda m_1 \phi(x_1)\langle \psi_x\rangle (x_1)+\lambda m_1\psi(x_1)\langle \phi_x\rangle (x_1)\right)\delta_{x_1}=2\lambda \langle(\phi \psi)_x\rangle \rho, 
\end{split} 
\end{equation*} 
which, in conjuncture with the first calculation and in view of equation \eqref{eq:weakL}, implies
\begin{equation*}
\frac12 D_x^3(\phi\psi)=\lambda \In (\phi\psi), 
\end{equation*}
which completes the proof.  

\end{proof}

\begin{lemma}\label{lem:b1solgenD}
	Let $\lambda \in \mathbb{C}$ and $\rho$ be the discrete measure defined by 
	\eqref{eq:drho}.  
	Moreover, let $\bb=\phi\psi$ where $\phi$ and $\psi$ satisfy
	\begin{equation*}
	\begin{split}	
		&D^2_x\phi=\lambda \rho\phi,\qquad \phi_{x}(0)-h\phi(0)=0,\\
		&D^2_x\psi=\lambda \rho\psi,\qquad \psi_{x}(1)+H\psi(1)=0.
	\end{split}
	\end{equation*}

	Then $\bb$ satisfies 
	\begin{equation}\label{eq:DSelfAdjointThird}
		\frac12 D^3_x \bb-\lambda\In \bb=0,
	\end{equation}
	with the boundary conditions \eqref{eq:sGenBC1}\rm{:}
	\begin{equation*}
	\begin{split}
		&\frac{1}{2}\bb_{xx}(0)-h\bb_{x}(0)+h^2\bb(0)=0,\\
		&\frac{1}{2}\bb_{xx}(1)+H\bb_{x}(1)+H^2\bb(1)=0.  
	\end{split}
	\end{equation*}
\end{lemma}
\begin{proof}
The only thing to check are the boundary conditions.  However, 
the support of $\rho$ is away from the end points so the computation takes 
place in the smooth sector and, as a consequence, the computation done 
previously applies.  
\end{proof}
Finally, we present appropriate analogs of Lemma \ref{lem:Dproduct} and Theorem \ref{thm:bconstruction}.  We recall that the function defining the spectrum of 
the boundary value problem \eqref{eq:stringBVP}  is given, in our notation, 
by $D(-\lambda)=\phi_x(1;\lambda)+H \phi(1;\lambda), 0\leq H$, where the eigenvalue $z$ in the original problem is taken to be $-\lambda$.  We also recall 
(see \eqref{eq:cn}, \eqref{eq:hatcn} ) that $c_0(x)=\phi(x;0), \, \hat c_0(x)=\psi(x;0)$.

\begin{lemma} \label{lem:DproductD}
Let $\rho$ be the discrete measure defined by 
	\eqref{eq:drho} and let $W(f,g)$ denote the 
Wronskian of two functions $f,g$.  Suppose 
that $h+H>0$ and let the eigenvalues of the string problem \eqref{eq:stringBVP} be denoted by $0<z_1<z_2<\dots<z_N$, then $D$ admits the additive representation
\begin{equation}\label{eq:DformulaD}
D(-\lambda)=W(\hat c_0,c_0)+\sum_{n=1} ^N\lambda^n 
\sum_{1\leq i_1<i_2<\dots i_n \leq N} 
m_{i_1}m_{i_2}\dots m_{i_n} 
(x_{i_n}-x_{i_{n-1}})(x_{i_{n-1}}-x_{i_{n-2}})\dots (x_{i_2}-x_{i_1}) c_0(x_{i_1})\hat c_0(x_{i_n}),  
\end{equation}
and the product representation
\begin{equation}\label{eq:DformulaprodD}
D(-\lambda)=W(\hat c_0,c_0)\prod_{j=1}^N \left(1+\frac{\lambda}{z_j}\right).  
\end{equation}
\end{lemma}
\begin{proof}
The additive formula follows from the definitions \eqref{eq:cn}, \eqref{eq:hatcn} of $c_n$, 
 $\hat c_n$, respectively.   
The multiplicative formula is well known and it follows immediately from 
Hadmard's product formula and the fact that the spectrum is finite 
because $m$ is a finite discrete measure.  
\end{proof} 
 \begin{theorem} \label{thm:bconstructionD}
Let $\rho$ be the discrete measure given by \eqref{eq:drho}, $\beta, K$ be defined as in Proposition \ref{prop:cond-isospectrality}, using equations \eqref{eq:betaleft}, \eqref{eq:Kequation}  respectively, and let $\omega$ 
be given as in Lemma \ref{lem:b1solgenD}. 
Then 
\begin{enumerate} 
\item $\phi, \psi$ defined in Lemma \ref{lem:b1solgenD} are polynomials of degree $N$;
\item $D(-\lambda)=\phi_x(1;\lambda)+H\phi(1;\lambda)=-\left(\psi_x(0;\lambda)-h\psi(0;\lambda)\right)$;
\item $b_{-1}, b_0$ defined by 
\begin{equation} \label{eq:b0b-1epsilon}
\begin{gathered}
b_{-1}(x;\epsilon)=\bb(x; \epsilon),\\
b_0(x;\epsilon)=\frac{\big[\bb(x;0)-\bb(x;\epsilon)\big]}{\epsilon},\\ 
, 
\end{gathered} 
\end{equation}
satisfy equation \eqref{eq:CGS1cons} with $\In$ defined by \eqref{eq:weakL}. 
\item   
$$\beta(z;\epsilon)=\frac{1}{2}\big[\frac{D(-\epsilon)-D(0)}{\epsilon}-
\frac{D(-\epsilon)}{z+\epsilon}\big], $$
\item 
$K=-1$, and $\beta(z;\epsilon)$ is deformation invariant ($\dot \beta=0$).  
\item Moreover, if $h+H>0$ and $0<z_1<z_2<\dots < z_N$ denotes the spectrum of the 
boundary value problem \eqref{eq:stringBVP} then 
\begin{equation} \label{eq:Depsilon}
D(-\epsilon)=W(\hat c_0,c_0)\prod_{j=1}^N\left(1+\frac{\epsilon}{z_j}\right).
\end{equation}  
\end{enumerate} 
\end{theorem} 
\begin{proof}
The assertions of the theorem follow from the fact that $c_n$ and $\hat c_n$ 
defined by equations \eqref{eq:cn} and \eqref{eq:hatcn} are valid if $\rho$ is 
any positive measure, in particular if $\rho$ is finite and discrete.  The series 
(in $\lambda$) which define $\phi$ and $\psi$ (see the proof of Lemma \ref{lem:sb1xGenRevised}) terminate at $n=N$ as can be easily verified.  
For example: the term $\prod_{j=1}^n(\xi_{j+1}-\xi_j)\rho(\xi_j)$ will  vanish identically on the domain $0\leq \xi_1\leq \xi_2\leq \dots \leq \xi_{n+1}=x$ if $n>N$. 
In particular $D(z)=\phi_x(1;-z)+H\phi(1;-z)$ is a polynomial of degree $N$ and 
$D(z)=-(\psi_x(0;-z)-h\psi(0;-z))$ by either directly inspecting iterates $c_n$ and $\hat c_n$ or by observing that $\phi_x(1;\lambda)+H\phi(1;\lambda)$ and $\psi_x(0;\lambda)-h\psi(0;\lambda)$ can only differ by a constant multiplier, independent of $\lambda$, which can easily be verified to be  $-1$ for $\lambda=0$.  
The statements about $\beta$ and $K$ then follow from equations \eqref{eq:betah}, 
\eqref{eq:betaphi0}, \eqref{eq:betaphi1}.  Finally, the invariance of the spectrum implies 
that $\beta$ is constant in $t$.  
 \end{proof}
 
 We recall the deformation equation \eqref{eq:Ddeformationeq}
 \begin{equation*}
z D_t\rho=\frac12 D^3_x b+z\In b. 
\end{equation*}
 Evaluating this equation on the support of $\rho$ with the help of 
 the previous theorem and the definition of $\In$ given by \eqref{eq:weakL} we obtain the following explicit 
 form of the evolution equation, which we state for simplicity in the simplest case of a single 
 pole rational flow.  
\begin{corollary} \label{cor:epsilonD}
Let the ZS deformation be given by $b=b_0+\frac{b_{-1} }{z+\epsilon}$, 
 $\rho=\sum_{j=1}^N m_j \delta_{x_j}, \,\,  0<x_1<\dots<x_N<1$,  
 and let $b_0$ and $b_{-1}^{(k)}$ be given by equations \eqref{eq:b0b-1epsilon}.  
 Then the positions $x_j$ and masses $m_j$ evolve according to: 
 \begin{center}
 \boxed{
 \dot x_j=-b_0(x_j), \qquad \dot m_j=m_j\langle b_{0,x}\rangle (x_j), \qquad 1\leq j\leq N.  }
 \end{center}
 The sums 
 \begin{center} 
 \boxed{I_1=\sum_{i=1}^N m_i c_0(x_i)\hat c_0(x_i), }
 \end{center} 
 and 
 \begin{center}
 \boxed{
 I_n=\sum_{1\leq i_1<i_2<\dots i_n \leq N} 
m_{i_1}m_{i_2}\dots m_{i_n} 
(x_{i_n}-x_{i_{n-1}})(x_{i_{n-1}}-x_{i_{n-2}})\dots (x_{i_2}-x_{i_1}) c_0(x_{i_1})\hat c_0(x_{i_n}), \quad  2\leq n\leq N}
\end{center} 
 are invariants of the ZS deformation (constants of motion).  
 \end{corollary}
 
 It is worth mentioning that the deformation parameter $t$ has no direct physical 
 meaning, and as such it can be chosen arbitrarily.  Since the constraints 
 on the fields $b_{-1}, b_0$ are homogenous, one can rescale 
 both fields by a common, $x$-independent factor.  
 The rescaling by the Wronskian $W(c_0,\hat c_0)$ is particularly natural as it allows one to formulate equations in terms of Green's functions.  
\begin{corollary} \label{cor:epsilonD2nd}
Let the ZS deformation be given by $b=b_0+\frac{b_{-1} }{z+\epsilon}$ with 
 $\rho=\sum_{j=1}^N m_j \delta_{x_j}, $ where $0<x_1<\dots<~x_N<1$, and let $h+H>0$
 in the boundary value problem \eqref{eq:stringBVP}.  
 Let $G(x,y;\lambda)$  be 
 Green's function of $D_x^2-\lambda \rho$ to boundary 
conditions $G_x(0,y)-hG(0,y)=0, ~G_x(1,y)+HG(1,y)=~0$
 and define new $b_0$ and $b_{-1}$ by: 
 \begin{equation*}
b_{-1}(x;\epsilon) \stackrel{\emph{def}}{=} G(x,x;\epsilon), \quad b_0(x;\epsilon)\stackrel{\emph{def}}{=}
\frac{1}{\epsilon}\left(G(x,x;0)-G(x,x;\epsilon) \right).  
\end{equation*}
Then the positions $x_j$ and masses $m_j$ evolve with respect to the rescaled time $\tilde t=W(c_0,\hat c_0)t$ according to: 
 \begin{center}
 \boxed{
 \dot x_j=-b_0(x_j), \qquad \dot m_j=m_j\langle b_{0,x}\rangle (x_j), \qquad 1\leq j\leq N,   }
 \end{center}
 and the sums 
 \begin{center} 
 \boxed{I_1=\sum_{i=1}^N m_i G(x_i,x_i; 0), }
 \end{center} 
 and 
 \begin{center}
 \boxed{
 I_n=\sum_{1\leq i_1<i_2<\dots i_n \leq N} 
m_{i_1}m_{i_2}\dots m_{i_n} 
(x_{i_n}-x_{i_{n-1}})(x_{i_{n-1}}-x_{i_{n-2}})\dots (x_{i_2}-x_{i_1}) G(x_{i_1},x_{i_n}; 0), \quad  2\leq n\leq N}
\end{center} 
 are invariants of the ZS deformation (constants of motion).  
 \end{corollary}

\subsection{Linearization of the spectral data}
Using the expression for $\beta$ from Theorems \ref{thm:bconstruction}, \ref{thm:bconstructionD} along with Proposition \ref{prop:cond-isospectrality}, in particular equation \eqref{eq:Kequation}, we find that the following statements can be made for general boundary conditions.  For brevity we will use  
the abbreviations $\mathcal{D}, \mathcal{N}, \mathcal{M} $ for Dirichlet, Neumann, or mixed boundary conditions respectively as well we will suppress, till further notice, the dependence on $x,z$ and $\epsilon$.

\begin{theorem} \label{thm:Big} 
Let the ZS deformation be given by $b=b_0+\frac{b_{-1} }{\epsilon +z},\, \epsilon>0$ with 
the mass density $\rho \in C^1([0,1])$ which vanishes in a neighbourhood of each endpoint, or $\rho=\sum_{j=1}^N m_j \delta_{x_j}, \qquad 0<x_1<\dots<x_N<1$.  
Then the string boundary value problem \eqref{eq:stringBVP} is isospectral with the mass density $\rho$ evolving according to
\begin{equation*}
\begin{split}
         &\rho_t=\In b_{0}, \\
         &\epsilon b_{0,xxx}+b_{-1,xxx}=0, \\
         &\frac12 b_{-1,xxx}-\epsilon \In b_{-1,x}=0,
\end{split}
\end{equation*}
with $b_0, b_1, \beta$ constructed in Theorems \ref{thm:bconstruction}, \ref{thm:bconstructionD} and $\In$ defined by equations \eqref{eq:Lrho}, \eqref{eq:weakL}.  
In addition, the evolution of the spectral data $A_1, B_1$ (equation \eqref{eq:vI1BC}) for general boundary conditions 
linearizes as follows: 
\begin{enumerate}
	\item $\mathcal{D}-\mathcal{D}$: the spectrum is given by $B_{1}=0$, the evolution by
		\begin{equation*}\label{eq:DirDir}
			\dot{A}_{1}=2\beta A_{1}-\frac{1}{2}b_{xx}(1)B_{1},\qquad \dot{B}_{1}=0.
		\end{equation*}
	\item $\mathcal{D}-\mathcal{N}$: the spectrum is given by $A_{1}=0$, the evolution by
		\begin{equation*}\label{eq:DirNeu}
			\dot{A}_{1}=0,\qquad \dot{B}_{1}=2\beta B_{1}+b(1)A_{1}.
		\end{equation*}
	\item $\mathcal{N}-\mathcal{N}$: the spectrum is given by $A_{1}=0$, the evolution by
		\begin{equation*}\label{eq:NeuNeu}
			\dot{A}_{1}=2\beta A_{1},\qquad \dot{B}_{1}=b(1)A_{1}.
		\end{equation*}
	\item $\mathcal{N}-\mathcal{D}$: the spectrum is given by $B_{1}=0$, the evolution by
		\begin{equation*}\label{eq:NeuDir}
			\dot{A}_{1}=2\beta A_{1}-\frac{1}{2}b_{xx}(1)B_{1},\qquad \dot{B}_{1}=0.
		\end{equation*}
	\item $\mathcal{D}-\mathcal{M}$: the spectrum is given by $A_{1}+HB_{1}=0$, the evolution by
		\begin{equation*}\label{eq:DirMix}
			\dot{A}_{1}+H\dot{B}_{1}=0, \qquad \dot{A}_{1}-H\dot{B}_{1}=2\beta[A_{1}-HB_{1}]+b_{x}(1)[A_{1}+HB_{1}].
		\end{equation*}
	\item $\mathcal{M}-\mathcal{D}$: the spectrum is given by $B_{1}=0$, the evolution by
		\begin{equation*}\label{eq:MixDir}
			\dot{A}_{1}=2\beta A_{1}-\frac{1}{2}b_{xx}(1)B_{1},\qquad \dot{B}_{1}=0.
		\end{equation*}
	\item $\mathcal{N}-\mathcal{M}$: the spectrum is given by $A_1+HB_{1}=0$, the evolution by
		\begin{equation*}\label{eq:NeuMix}
			\dot{A}_{1}+H\dot{B}_{1}=0, \qquad \dot{A}_{1}-H\dot{B}_{1}=2\beta[A_{1}-HB_{1}]+b_{x}(1)[A_{1}+HB_{1}].
		\end{equation*}
	\item $\mathcal{M}-\mathcal{N}$: the spectrum is given by $A_{1}=0$, the evolution by
		\begin{equation*}\label{eq:MixNeu}
			\dot{A}_{1}=0,\qquad \dot{B}_{1}=2\beta B_{1}+b(1)A_{1}.
		\end{equation*}
	\item $\mathcal{M}-\mathcal{M}$: the spectrum is given by $A_1+HB_{1}=0$, the evolution by
		\begin{equation*}\label{eq:MixMix}
			\dot{A}_{1}+H\dot{B}_{1}=0, \qquad \dot{A}_{1}-H\dot{B}_{1}=2\beta[A_{1}-HB_{1}]+b_{x}(1)[A_{1}+HB_{1}].
		\end{equation*}
\end{enumerate}
\end{theorem} 
\begin{remark}
	Observe that even though the same type of boundary conditions on the right end of the string formally yields the same flow equations of the spectral data, 
the coefficient $\beta$ is sensitive to the boundary conditions on the left end of the string, and so is the field $b$.
\end{remark}

\begin{remark} Theorem \ref{thm:Big} can be readily generalized to the case 
of multi-pole rational flows \eqref{eq:rationalD} along the lines of Theorem \ref{thm:bratconstruction}.
\end{remark} 
\subsection{ The limit $\epsilon \rightarrow 0$}
In previous sections we constructed the fields $b_{-1}(x;\epsilon), b_0(x;\epsilon)$ in 
such a way that both are entire functions of $\epsilon$.  We recall equations that 
$\rho$ and $b$ satisfy: 
\begin{equation*} 
    \rho_{t}=\In b_0, \qquad 
    b_{0,xxx}+\frac{b_{-1,xxx}}{\epsilon}=0, \qquad 
    \frac{1}{2}b_{-1,xxx}^{} -\epsilon\In b_{-1}=0, 
\end{equation*} 
which we write as
 \begin{equation*} 
    \rho_{t}=\In b_0, \qquad 
  \frac12 b_{0,xxx}+\In b_{-1}=0, \qquad 
    \frac{1}{2}b_{-1,xxx}^{} -\epsilon\In b_{-1}=0.   
\end{equation*} 
We now take  the limit $\epsilon \rightarrow 0$.  Without giving a detailed account of intermediate elementary steps we summarize the results
 in the 
following theorem.  
\begin{theorem} \label{thm:gCH} Let the ZS deformation be given by $b=b_0+\frac{b_{-1} }{\epsilon +z},\, \epsilon>0$ with 
the mass density $\rho \in C^1([0,1])$ which vanishes in a neighbourhood of each endpoint, or $\rho=\sum_{j=1}^N m_j \delta_{x_j}, \,0<x_1<\dots<x_N<1$.  
Let $b_{-1}(x;\epsilon)$ and $b_0(x;\epsilon)$ be given as in theorems \ref{thm:bconstruction} and \ref{thm:bconstructionD}.  
Then 
\begin{enumerate} 
\item 
\begin{equation*}
    \begin{gathered} 
    \lim_{\epsilon \rightarrow 0} b_{-1}(x;\epsilon)=c_0(x)\hat c_0(x), \\
    \lim_{\epsilon \to 0} b_0(x;\epsilon)=-(c_0(x)\hat c_1(x)+c_1(x) \hat c_0(x)), \\
    D(0)=W(\hat c_0,c_0), \qquad 
    D'(0)=\int_0^1 \hat c_0(\xi)\rho(\xi) c_0(\xi) d\, \xi,   
    \end{gathered} 
\end{equation*}
where $W(f,g)$ denotes the Wronskian of $f,g$.  
\item If in addition $h+H>0$ and the fields $b_0$ and $b_{-1}$ are redefined as: 
\begin{equation} \label{eq:b0b-1}
\begin{gathered}
b_{-1}(x)\stackrel{\text{def}}{=}\frac{c_0(x)\hat c_0(x)}{W(c_0,\hat c_0)} , \\\\
b_0(x)\stackrel{\text{def}}{=} \, \frac{-(c_0(x)\hat c_1(x)+c_1(x) \hat c_0(x))}{W(c_0, \hat c_0)}, 
\end{gathered}
\end{equation}
 
then 
\begin{center}
\boxed{ 
b_{-1}(x)=G(x,x), \quad b_0(x)=-\int _0^1 \abs{x-\xi} G(x,\xi) \rho(\xi) d\xi}
\end{center} 
where $G(x,y)$ is Green's function of $D_x^2$ to boundary 
conditions $G_x(0,y)-hG(0,y)=0, ~G_x(1,y)+HG(1,y)=~0$
and $\beta(z)$ (see Theorem \ref{thm:Big}) satisfies 
\begin{center}
\boxed{\beta(z)=-\frac12\left(\int_0^1 G(\xi,\xi)\rho(\xi)d\xi -\frac1z\right)}
\end{center} 

\item With $b_0, b_{-1}$ so defined $\rho$ undergoes an isospectral deformation governed by
\begin{equation*} 
    \rho_{\tilde t}=\In b_0, \qquad 
  \frac12 b_{0,xxx}+\In b_{-1}=0, \qquad 
    b_{-1,xxx}=0, \quad \textrm{ where } \tilde t=W(c_0,\hat c_0)t.    
\end{equation*} 

\end{enumerate} 
\end{theorem} 
\begin{remark} 
Dividing by the Wronskian, which is constant, is predicated on the absence 
of $0$ in the spectrum of the original problem.  For Neumann-Neumann 
boundary conditions $0$ is an eigenvalue and $W(c_0,\hat c_0)=0$.  
Then one possible choice is $c_0=\hat c_0=\frac{1}{\sqrt{2}}$ and $b_{-1}=\frac12, b_0(x)=\int_0^1G_T(x,\xi)\rho(\xi)d\xi$, where $G_T(x,y)=-\frac{\abs{x-y}}{2}$ is the translationally invariant Green's function of 
$-D^2_x$.  
\end{remark} 
We will finish this section stating explicit formulas 
for the evolution of positions $x_i$ and weights $m_i$ in the case of the discrete measure $\rho$.  
These formulas are implicit in the statement of the previous theorem and we present them only to emphasize their simplicity.   
\begin{corollary} \label{cor:gCH}
Let the ZS deformation be given by $b=b_0+\frac{b_{-1} }{z},$ with 
 $\rho=\sum_{j=1}^N m_j \delta_{x_j}, \qquad 0<x_1<\dots<x_N<1$ 
 and let $b_0$ and $b_{-1}$ be given by equations \eqref{eq:b0b-1}.  
 Then the positions $x_j$ and masses $m_j$ evolve with respect to the rescaled time $\tilde t=W(c_0,\hat c_0)t$ according to: 
 \begin{center}
 \boxed{
 \dot x_i=-b_0(x_i), \qquad \dot m_i=m_i\langle b_{0,x}\rangle (x_i), \qquad 1\leq i\leq N,  }
 \end{center}
 and the sums 
 \begin{center} 
 \boxed{I_1=\sum_{i=1}^N m_i G(x_i,x_i), }
 \end{center} 
 and 
 \begin{center}
 \boxed{
 I_n=\sum_{1\leq i_1<i_2<\dots i_n \leq N} 
m_{i_1}m_{i_2}\dots m_{i_n} 
(x_{i_n}-x_{i_{n-1}})(x_{i_{n-1}}-x_{i_{n-2}})\dots (x_{i_2}-x_{i_1}) G(x_{i_1},x_{i_n}), \quad  2\leq n\leq N}
\end{center} 
 are invariants of the ZS deformation (constants of motion).  
 \end{corollary}
\section{Examples of isospectral deformations of a discrete string}
\label{sec:negativeflows}

\begin{example} 
As an example of the developed formalism we consider the case of the string boundary value problem \eqref{eq:stringBVP} corresponding to the mass density $\rho=~\sum_{j=1}^{N}m_{j}\delta_{x_j}, \, 0<x_1<x_2< \dots x_n<1$, with boundary conditions  of the type $0<h<\infty$ and $H=0$. Let
\begin{equation*}
	v|_{I_{j}}=v_{j}=p_{j}(x-x_{j})+q_{j},\qquad \text{where } I_{j}=(x_{j},x_{j+1}),
\end{equation*}
denote the solution to the initial value problem $-v_{xx}=z\rho v, \,  v_x(0)-hv(0)=0$
with the convention that $x_{0}=0$ and $x_{N+1}=1$.  The construction of this solution proceed as follows.  

From the boundary condition at $x=0$, one can take $p_{0}=h$ and $q_{0}=1$.  Letting $l_{j}$ denote the length of the interval $I_j$ and imposing continuity of $v$ at $x=x_{j+1}$ one finds that $p_{j}=\frac{q_{j+1}-q_{j}}{l_{j}}$.  The jump in the derivative of $v$ at $x=x_{j+1}$ results in $p_{j+1}-p_{j}=-zm_{j+1}q_{j+1}$. On the last interval $I_N$  $v_{N}=A_{1}(x-1)+B_{1}=p_{N}(x-x_{N})+q_{N}$ from which one finds that $A_{1}=p_{N}$ and $B_{1}=p_{N}l_{N}+q_{N}$.  Let us define the Weyl function for this problem 
\begin{equation*}
	W(z)\stackrel{\emph{def}}{=}\frac{v(1;z)}{v_{x}(1;z)}=\frac{B_{1}}{A_{1}}. 
\end{equation*}
From the construction $W(z)=\frac{p_Nl_N+q_N}{p_N}=l_N +\frac{q_N}{p_N}$. 
Iterating with the help of continuity and jump conditions we obtain 
\begin{equation*}
W(z)=l_{N}+\cfrac{1}{-zm_N+\cfrac{1}{l_{N-1}+\cfrac{1}{\ddots+\cfrac{1}{l_{0}+\frac{1}{h}}}}}. 
\end{equation*}

Since the spectrum of the boundary value problem at hand is simple (only simple eigenvalues occur) we have the following partial fraction decomposition: $W(z)=W_{\infty}+\sum_{i=1}^{N}\frac{a_{i}}{-z+z_{i}}$, where $W_{\infty}=l_{N}$ and
\begin{equation*}
	a_{i}=-\lim_{z\rightarrow z_{i}}\frac{B_{1}(z)}{A_{1}(z)}(z-z_{i})=-\frac{B_{1}(z_{i})}{\frac{dA_{1}}{dz}|_{z=z_{i}}}.
\end{equation*}
It can be shown that the residues $a_i$ are strictly positive.  We recall that 
the spectrum $\{z_j: 1\leq N\}$ is determined from $A_1(z)=0$.  Following Theorem \ref{thm:Big}, item (8), we get $\dot{A}_{1}=0$, and $\dot{B}_{1}(z_{i},t)=2\beta(z_{i})B_{1}(z_{i},t)$  so that
\begin{equation*}
	B_{1}(z_{i},t)=B_{1}(z_{i},t=0)e^{2\beta(z_{i})t}.  
\end{equation*}

This implies that the residues evolve as \begin{equation*}
	a_{i}(t)=a_{i}(0)e^{2\beta(z_{i})t}.
\end{equation*}
Thus the Weyl function can be written as
\begin{equation*}
	W(z,t)=l_{N}+\sum_{i=1}^{N}\frac{a_{i}(0)e^{2\beta(z_{i})t}}{-z+z_{i}}\stackrel{\emph{def}}{=}l_N +\int \frac{e^{2\beta(\zeta)t}d\mu(\zeta)}{-z+\zeta}
	\end{equation*}
where $d\mu(\zeta)=\sum_{i=1}^N a_i(0) \delta _{z_i}$ is the spectral measure 
at $t=0$.  	Hence if we set $\lambda=-z$ 
we obtain the formula 
\begin{equation*}
\int \frac{e^{2\beta(\zeta)t}d\mu(\zeta)}{\lambda+\zeta}=\cfrac{1}{\lambda m_N+\cfrac{1}{l_{N-1}+\cfrac{1}{\ddots+\cfrac{1}{l_{0}+\frac{1}{h}}}}}, 
\end{equation*}
which shows that using Stieltjes' inversion formulas (see \cite{Stieltjes, BSS-Stieltjes}) we can recover $m_N, \cdots, m_1$ and $l_{N-1}, \cdots, l_0$ in 
terms of the moments of the measure $e^{2\beta(\zeta)t} d\mu(\zeta)$.  
Finally, one can recover $l_N$ by observing that 
\begin{equation*}
W(0)=\frac{h+1}{h}=1+\frac1h=l_N+\int \frac{e^{2\beta(\zeta)t}d\mu(\zeta)}{\zeta}
\end{equation*}
and solve for $l_N$, or compute $l_N$ from the formula for the 
total length of the string: $l_N=1-\sum_{i=0}^{N-1}l_i$.  The formula for 
$\beta(\zeta)$ depends on the deformation and boundary conditions and is given 
by \eqref{eq:b0b-1epsilon}, \eqref{eq:b0b-1}, if $\epsilon>0$, $\epsilon=0$ respectively.  More generally, for the multi-pole model, $\beta(\zeta)$ 
is given by \eqref{eq:betarat}.

\end{example} 

\begin{example} \label{ex:DD/DN}
Let us now consider the case of $\epsilon =0$, hence $b=b_0+\frac{b_{-1}}{z}$, and let us assume 
a discrete mass density $\rho=\sum_{j=1}^N m_j \delta_{x_j}$.  
For the sake of comparison we consider two cases of 
boundary conditions: Dirichlet-Dirichlet and Dirichlet-Neumann.  
We point out that the Dirichlet-Neumann case appeared, 
somewhat unexpectedly, in the recent work on the modified Camassa-Holm equation
\cite{ChangS}.

We use the formulas corresponding to the rescaled time (see 
theorem \ref{thm:gCH} and corollary \ref{cor:gCH}).

For the Dirichlet-Neumann case we have 
\begin{equation*}
    \begin{gathered}
    G^{DN}(x,y)=-\begin{cases}
    x, \quad x<y\\ y, \quad x>y 
    \end{cases}\\
    b_{-1}=-x, \qquad b_0=-\int_0^1 \abs{x-\xi} G^{DN}(x, \xi) \rho(\xi) d\xi=\sum_{x_j< x}(x-x_j)x_j m_j+\sum_{x_j>x} (x_j-x)x m_j, \\
    \end{gathered}   
\end{equation*}
Both $b_{-1}, b_0$ satisfy the required boundary conditions $b(0)=0,\, b_{xx}(1)=0$.

For the Dirichlet-Dirichlet case we have 
\begin{equation*}
    \begin{gathered}
    G^{DD}(x,y)=-\begin{cases}
    x(1-y), \quad x<y\\ y(1-x), \quad x>y 
    \end{cases}\\
    b_{-1}=-x(1-x), \qquad b_0=-\int_0^1 \abs{x-\xi} G^{DD}(x, \xi) \rho(\xi) d\xi=\sum_{x_j<x}(x-x_j)(1-x)x_j m_j+\sum_{x_j>x} (x_j-x)x (1-x_j)m_j, \\
    \end{gathered}   
\end{equation*}
Clearly, $b_{-1}, b_0$ satisfy the required boundary conditions $b(0)=0,\, b(1)=0$. 

In both cases, the explicit form of the evolution equations (with respect to the rescaled time) reads: 
\begin{align*}
\dot x_i&=\sum_{j=1}^N \abs{x_i-x_j} G(x_i,x_j)m_j, \\
\dot m_i&=-m_i \left(\sum_{j=1}^N \sgn{(x_i-x_j)} G(x_i,x_j)m_j+\sum_{j=1}^N 
\abs{x_i-x_j} \langle G_x(x,x_j)\rangle(x_i) m_j\right), \qquad 
\sgn(0)=0. 
\end{align*}
with $G$ being an appropriate Green's function.   
\end{example}

\section{Acknowledgments}~
The authors would like to thank Richard Beals and Xiangke Chang for helpful comments on earlier drafts of the paper, and an anonymous referee for 
making penetrating comments and for asking questions which led to an essential 
improvement of the quality of this paper.  

This work was supported by  the Natural Sciences and Engineering Research Council of Canada
[NSERC USRA 220082 to K. Colville, NSERC USRA 414987 to D. Gomez  and NSERC 163953 to J. Szmigielski] and by the Department of Mathematics and 
Statistics of the University of Saskatchewan.  

\section{Appendix I: weak Lax pairs}\label{app:weak formulation}
We assume $\rho=\sum_{j=1}^N m_j \delta_{x_j}$ for the remainder of the appendix.  
Our goal is to compute conditions on the Lax pair to be compatible as a distributional Lax pair.  
We will denote by $D_x, D_t$ the distributional derivatives with respect to 
$x,t$ respectively, and define the distributional Lax pair as: 
\begin{equation}\label{eq:DxLax}
D_x V=\begin{bmatrix}0&1\\-z \rho&0\end{bmatrix}V \qquad  D_tV=\begin{bmatrix}-\frac12 D_x(b)+\beta& b\\-\frac12 D^2_x(b)-z\rho b& \frac12 D_x(b) +\beta \end{bmatrix}V.   
\end{equation}
It is immediate from the first equation that the first component of $V$ is 
continuous and piecewise smooth with jumps only on the support of $\rho$.  
So we can write, as before, $V=\begin{bmatrix} v\\v_x \end{bmatrix}$, where $v_x$ is the classical 
derivative.  This implies that the first member of the Lax pair contains only one 
non-trivial statement: $D_x(v_x)=D^2_x(v)=-z\rho v $ which is equivalent to 
\begin{equation} \label{eq:vxjump}
v_{xx}=0, \quad  \text{ if } x\neq x_j, \qquad [v_x](x_j)=-zm_j v(x_j), 
\end{equation}
where $[f](x_j)$ means the jump of $f$ at $x_j$.  
Likewise, the second member of the Lax pair is well defined if 
$b$ is continuous and piecewise smooth with jumps only on the support of $\rho$.  
Thus the second equation can be written: 
\begin{equation}\label{eq:DtLax}
D_tV=\begin{bmatrix} v_t\\ v_{tx}-\sum_{i=1}^N [v_x](x_i) \dot x_i \delta_{x_i} \end{bmatrix}=\begin{bmatrix}-\frac12 D_x(b)+\beta& b\\-\frac12 D^2_x(b)-z\rho b& \frac12 D_x(b) +\beta \end{bmatrix}V,  
\end{equation}
or, equivalently, 
\begin{equation*}
\begin{gathered} 
v_t=(-\frac12 b_x+\beta)v+bv_x, \\
 v_{tx}-\sum_{i=1}^N [v_x](x_i) \dot x_i \delta_{x_i} =-\frac12 b_{xx}v-\sum_{j=1}^N\left(\frac12 [b_x](x_j)v(x_j)+zm_jb(x_j)v(x_j)\right) \delta_{x_j}+(\frac12b_x+\beta)v_x, 
 \end{gathered}
 \end{equation*}
showing that it is the singular part of $D_tV$ that determines the evolution of $x_i$; using 
 \eqref{eq:vxjump} and \eqref{eq:DtLax} we obtain
 \begin{equation} \label{eq:xdot}
 zm_i\dot x_i=-\left(\frac12[b_x](x_i)+zm_i b(x_i)\right).  
\end{equation}
The actual compatibility condition of equations \eqref{eq:DxLax}, \eqref{eq:DtLax}, 
which can be succinctly formulated as distributional equality $D_xD_tV=D_tD_xV$, is presented in the 
following theorem.  
\begin{theorem} Let $\rho=\sum_{j=1}^N m_j \delta_{x_j}$ and let $b$ be 
continuous, piecewise differentiable function with jumps only on the support of $\rho$.   
Then $D_xD_tV=D_tD_x V$ implies: 
\begin{equation}
zD_t\rho=\frac12 D_x^3 b+z\In b, 
\end{equation} 
where $\In b=D_x (\rho b)+\langle b_x \rangle \rho$ and $ \langle b_x \rangle $ 
means the pointwise, arithmetic average of the right hand and left hand limits.  
\end{theorem} 
\begin{proof}
First we observe that the left hand side contains only the singular part of the distributional equality.  Indeed, 
\begin{equation*}
D_t \rho =\sum^N_{i=1}\dot m_i \delta_{x_i} -\sum_{i=1}^N m_i \dot x_i \delta_{x_i}'. 
\end{equation*}
Hence the regular part of the right hand side must be zero.  
Equating coefficients at $\delta _{x_j}$ and $\delta_{x_j}'$ on both sides we obtain 
that the statement of the theorem is equivalent to: 

\begin{subequations}
\begin{align}
b_{xxx}&=0, \qquad  \text {if }  \quad x\neq x_i, \\
z\dot m_i &=\frac12 [b_{xx}](x_i)+zm_i \langle b_x\rangle(x_i) \label{eq:zmdot}\\
-z m_i \dot x_i&=\frac12[b_x](x_i)+zm_ib(x_i) \label{eq:zxdot}, 
\end{align}
for $i=1,\dots N$.  
\end{subequations}
Moreover by localizing tests functions around individual 
points $x_i$ we can, without loss of generality, assume 
that $N=1$; in other words we can work with $\rho =m_1 \delta_{x_1}$.  
 
We obtain two statements corresponding to components of $D_tD_xV=D_xD_tV$:
\begin{subequations}
\begin{align}
&v_{tx}-\dot x_1 [v_x](x_1)\delta_{x_j}=\left((-\frac12 b_x+\beta)v+bv_x\right)_x +[-\frac12 b_x v +bv_x](x_1) \delta_{x_1}\\
\notag\\
&-z(\dot m_1 v(x_1)+m_1 \dot v(x_1))\delta _{x_1}+zm_1v(x_1) \dot x_1 \delta_{x_1}'=\\
&\left(-\frac12 b_{xx} v +\frac12 b_x v_x+\beta v_x\right)_x+[-\frac12 b_{xx} v+\frac12 b_x v_x +\beta v_x](x_1)\delta_{x_1}-\left(\frac12[b_x](x_1)v(x_1) +zm_1b(x_1) v(x_1)\right) \delta_{x_1}' \notag. \end{align}
\end{subequations}
The equality of regular parts implies: 
\begin{equation*}
v_{tx}=\left((-\frac12 b_x+\beta)v+bv_x\right)_x=v_{xt}, \quad b_{xxx}=0,  
\end{equation*}
while equating in the first equation the coefficients at $\delta_{x_1}$, with the help of \eqref{eq:vxjump}, one easily recovers \eqref{eq:zxdot}.  Finally, to 
prove \eqref{eq:zmdot} we need to equate the coefficients of $\delta_{x_1}$: 
\begin{equation*} 
-z(\dot m_1 v(x_1)+m_1 \dot v (x_1))=[-\frac12 b_{xx} v +\frac12 b_x v_x +\beta v_x](x_1),
\end{equation*}
and proceed by observing that 
\begin{enumerate} 
\item $\dot v(x_1)=\langle v_x \rangle \dot x_1+\langle v_t \rangle(x_1)$, 
\item $[b_x v_x]=\langle b_x \rangle  [v_x]+ [b_x]\langle v_x \rangle$. 
\end{enumerate}
Upon carrying out several elementary cancellations one arrives, somewhat exhausted, at the final formula \eqref{eq:zmdot}.  
\end{proof}
\section{Appendix II: mapping to the Real Line}\label{ap:mapping to R}
In this appendix we discuss a relation between the flow 
generated by $b=b_0+\frac{b_{-1}}{z}$ and the CH equation\cite{CH}(see 
also \cite{BSS-acoustic}), which we will 
write in the notation adapted to our setup: 
\begin{equation}\label{eq:CHm}
m_t=\Inm u, \qquad m=\frac12(u_{\zeta \zeta}-u),  \quad \zeta \in \R, 
\end{equation}
where $\Inm =m\frac{d}{d\zeta}+\frac{d}{d\zeta}m$.  We will discuss the relation 
for the smooth case only, although the comparison is also valid on the level of 
distributional Lax pairs.  
We recall the Lax pair formulation of CH:
\begin{equation}\label{eq:CHLax}
-\Psi_{\zeta \zeta}+\frac14 \Psi=zm \Psi, \qquad \Psi_t=(u-\frac{1}{z})\Psi_{\zeta}-\frac{u_{\zeta}}{2} \Psi, 
\end{equation}
whose compatibility condition gives the evolution equation on $m$ given 
above and the constraint 
\begin{equation}\label{eq:mconstraint}
m_{\zeta}=\frac12(u_{\zeta \zeta}-u)_{\zeta}.
\end{equation}

For us the starting equations are \eqref{eq:CH}
\begin{equation*}
	\rho_t = \In b_0,\qquad \dfrac{1}{2}b_{0,xxx} + \In b_{-1} = 0,\qquad b_{-1,xxx} = 0, 
\end{equation*}
where $\rho$ is the mass density in the boundary value problem undergoing 
an isospectral deformation
\begin{equation}\label{eq:stringLax}
-v_{xx}=z\rho v, \quad  \qquad 
v_t=(-\frac12 b_x+\beta)v+bv_x. 
\end{equation}

Now we perform a Liouville transformation:
\begin{equation}\label{eq:Liouville}
x = \frac{1}{2} + \frac{1}{2}\tanh{\frac\zeta2}, \qquad \Psi=(\cosh\frac\zeta2)\,  v, 
\end{equation}
which leads to an equivalent form of Lax equations \eqref{eq:stringLax}
on the real axis: 
\begin{equation}\label{eq:stringCHLax}
-\Psi_{\zeta \zeta}+\frac14 \Psi=z\big(\frac{1}{16} (\sech^4 \frac\zeta2)\rho\big) \Psi, \qquad \Psi_t=4(\cosh^2\frac\zeta2)b\Psi_{\zeta}+\big(-\frac12((4\cosh^2\frac\zeta 2)b)_{\zeta} +\beta\big)\Psi, 
\end{equation}
If we define: 
\begin{equation}\label{eq:uhats}
\hat u=4(\cosh^2\frac\zeta2)b\stackrel{def}{=}\hat u_0+\frac{\hat u_{-1}}{z}, \qquad \hat m=\frac{1}{16} (\sech^4 \frac\zeta2)\rho
\end{equation}
then the compatibility condition for \eqref{eq:stringCHLax} reads: 
\begin{subequations}
\begin{align}
&\hat m_t=\Inm \hat u_0, \\
&\frac12\left(\hat u_0-\hat u_{0,\zeta \zeta}\right)_\zeta=\Inm \hat u_{-1}, \label{eq:hatu0}\\
&\left(\hat u_{-1}-\hat u_{-1,\zeta \zeta}\right)_\zeta=0. 
\end{align}
\end{subequations}
For the Dirichlet-Dirichlet case (see Example \ref{ex:DD/DN}) 
$$b_{-1}=-x(1-x)=-\frac{1}{4(\cosh^2\frac\zeta2)}$$ which implies 
$\hat u_{-1}=-1$, which by equation \eqref{eq:hatu0} 
implies the constraint \eqref{eq:mconstraint}, and one recovers the CH case.  On the other hand, for example, for the Dirichlet-Neumann case $$b_{-1}=-x$$ and hence $$\hat u_{-1}=
-(e^{\zeta}+1)$$ which does not reduce to the CH case.

\bibliographystyle{abbrv}
\bibliography{NSERC2012}

\end{document}